\documentclass[12pt]{article}
\usepackage[top=1.18in, left=1.1in, right=1.1in, bottom=1in, includefoot]{geometry}
\usepackage{array}
\usepackage{booktabs}
\usepackage{rotating}
\usepackage{amsmath}
\usepackage{graphicx}

\usepackage{graphicx,url}

\usepackage{verbatim}

\usepackage{undertilde}
\usepackage[T1]{fontenc}
\usepackage{titlesec,blindtext}
\usepackage{enumitem}
\usepackage{graphicx}
\usepackage{placeins}
\usepackage{setspace}
 \usepackage{natbib}
\usepackage{listings}
\usepackage{datetime}
\usepackage{amssymb}

\usepackage{tabularx,colortbl}
\definecolor{Gray}{gray}{0.9}
\usepackage{booktabs}

\usepackage{color}

\newtheorem{thm}{Theorem}
\newtheorem{cor}{Corollary}
\newtheorem{conject}{Conjecture}
\titleformat{\section}
  {\large\bfseries}{\thesection}{1em}{}
\titleformat{\subsection}[block]{\hspace{1.5em}\normalsize\bfseries}{\thesubsection}{1em}{}
\titleformat{\subsubsection}[block]{\hspace{2.5em}\normalsize\bfseries}{\thesubsubsection}{1em}{}
\titleformat{\paragraph}[block]{\hspace{3.5em}\normalfont\normalsize\bfseries}{\theparagraph}{1em}{}

\begin{document}

\begingroup  
\centering
\large \textbf{Longitudinal Functional Data Analysis}\\[1em]
\large SY Park and AM Staicu \\ [0.5em]
\normalsize Department of Statistics, North Carolina State University\\ [0.2em]
\small spark13@ncsu.edu and astaicu@ncsu.edu\\[1em]
\today \\[1em]
\endgroup

\begin{abstract}
We consider analysis of dependent functional data that are correlated because of a longitudinal-based design: each subject is
observed at repeated time visits and for each visit we record a functional variable. We propose a novel parsimonious modeling framework for the repeatedly observed functional
variables that allows to extract low dimensional features. The proposed methodology
accounts for the longitudinal design, is designed for the study of the dynamic behavior of the underlying
process, and is computationally fast. Theoretical properties of this
framework are studied and numerical investigation confirms excellent behavior in finite samples. The proposed method is motivated by
and applied to a diffusion tensor imaging study of multiple sclerosis. Using Shiny \citep{shiny} we implement interactive plots to help visualize longitudinal functional data as well as the various components and prediction obtained using the proposed method. \\


\end{abstract}

\section{Introduction} \label{sec: introduction}

Longitudinal functional data refer to data consisting of functions (such as profiles or images) observed repeatedly for each subject at multiple instances, often visit times. Examples of such data include the Baltimore Longitudinal Study of Aging (BLSA), where daily physical activity count profiles are observed for each subject at several consecutive days \citep{xiao2013fast, goldsmith2014generalized}; and the longitudinal diffusion tensor imaging (DTI) study, where modality profiles along well-identified tracts are observed for each multiple sclerosis (MS) patient at several hospital visits \citep{greven2010}. As a result of an increasing number of such data, \textit{longitudinal functional data analysis} has received much attention recently; see for example \citet{morris2003wavelet, morris2006wavelet, baladandayuthapani2008bayesian, di2009multilevel, greven2010, staicu2010fast, chen2013repeated, li2014fpca}.  

Our motivation comes from the longitudinal DTI study, where the objective is to study the evolution of the MS disease as given by the dynamics of the modality profiles along the corpus callosum tract of the brain. By ``modality profile'' we mean measurements of a particular type of water diffusivity characteristics that are recorded at a fine grid of locations along the tract; the modality we focus on here is fractional anisotropy (FA). The change in the FA profiles along the corpus callosum over hospital visits is informative of the progression of the MS disease, and thus a statistical model for FA profiles, which incorporates longitudinal dependence, has the potential to be a very useful tool in practice. In this paper we propose a modeling framework that captures the process dynamics over time and also provides prediction of a full FA trajectory at a future visit. 

Currently available methods for analysis of longitudinal functional data mainly separate into two categories, based on whether or not they account for the actual time of subject's repeated visit, say $T_{ij}$ for the $j$th visit of subject $i$. However, most existing methods, including those that incorporate $T_{ij}$, cannot be used to predict a complete response trajectory at an unobserved (future) visit. For example, while \citet{greven2010} used a functional linear mixed model framework to model the process dynamics linearly in $T_{ij}$, their approach is not able to provide prediction of a full trajectory at a future visit. To the best of our knowledge, the only available modeling approach that provides such prediction is \citet{chen2013repeated}. Specifically, their proposed model, henceforth denoted CM, is $Y_{ij}(s) = \mu(s|T_{ij})+\sum_{k \geq 1} \xi_{ik}(T_{ij}) \phi_{k}(s|T_{ij})$, where 
$Y_{ij}(\cdot)$ is the $j$th repeated trajectory response measured at time $T_{ij}$, $\mu(\cdot|T_{ij})$ is the mean response function at $T_{ij}$, $\phi_{k}(\cdot|T_{ij})$ is the $k$th leading direction at $T_{ij}$, and $\xi_{ik}(T_{ij}) = \int \{ Y_{ij}(s| T_{ij})-\mu(s|T_{ij}) \}\phi_k(s|T_{ij}) ds $ is the corresponding  coefficient, which depends on $T_{ij}$. Though providing future prediction, CM makes the understanding of the process dynamics over visit times, $T_{ij}$, challenging. Additionally, the numerical investigation of CM in finite samples is restricted to 
the case when the sampling design for $T$'s, at which the functional response is observed, is dense or moderately sparse, as well as when the response curves are observed with white noise measurement error that has small magnitude relative to the variation of the response (\cite{chen2013repeated}). However, an important limitation of CM is its heavy computational cost; its rigorous study in diverse practical situations is almost impossible.

In this paper we focus on the case where the sampling design of $T$'s is sparse (hence \emph{sparse longitudinal design}) and the sampling design of $s$'s is dense (hence \emph{dense functional design}). We provide a novel parsimonious modeling framework to (i) easily study the longitudinal dynamics of the functional response over visit times, and (ii) predict a full trajectory at a future visit. We propose the following model for the response curve $Y_{ij}(\cdot)$:
\begin{equation}\label{eq: modeling framework}
Y_{ij}(s) = \mu(s,T_{ij}) + X_{i}(s, T_{ij}) + \epsilon_{ij}(s); \;\;  X_{i}(s, T_{ij})= \sum_{k} \xi_{ijk}\phi_k(s) 
\end{equation}
for $s \in\mathcal{S}$ and $T_{ij} \in \mathcal{T}$, where $\mu(\cdot, T_{ij})$ is an unknown smooth mean response corresponding to $T_{ij}$, $ X_i(\cdot, T_{ij})$ is a smooth random deviation from the mean that varies over time $T_{ij}$, and $\epsilon_{ij}(\cdot)$ is an error process with zero-mean and unknown covariance function. We assume that the bivariate processes $ X_i$'s are independent and identically distributed (iid), the error processes $\epsilon_{ij}$'s are iid and furthermore are independent of $X_i$'s. For identifiability we require that $X_{i}$ comprises solely the random deviation that is specific to the subject; 
any repeated time-specific deviation is viewed as part of $\epsilon_{ij}$. Here $\{\phi_k(s)\}_k$ are orthogonal basis functions in $L^2[0,1]$ and $\xi_{ijk} = \xi_{ik}(T_{ij})$ are the corresponding basis coefficients that have zero-mean, are uncorrelated over $i$, but correlated over $j$; the notation, $\xi_{ik}(T_{ij})$, is to show that this correlation may depend on $T_{ij}$. We assume that the set of visit times of all subjects, $\{T_{ij}:i, j\}$, is dense in $\mathcal{T}$. Full model assumptions are given in Section \ref{sec: model}.

The class of model (\ref{eq: modeling framework}) is rich and includes many existent models, as we now illustrate. (i) If $\xi_{ik}(T_{ij}) = \zeta_{0,ik} + T_{ij}\zeta_{1,ik}$ for appropriately defined random terms $\zeta_{0,ik}$ and $\zeta_{1,ik}$, model (\ref{eq: modeling framework}) can be represented as in \citet{greven2010}. (ii) If $\text{cov}(\xi_{ik}(T), \xi_{ik}(T') ) = \lambda_k \rho_k(|T-T'|; \nu)$ for some unknown variance $\lambda_k$, known correlation function $\rho_k(\cdot; \nu)$ with unknown parameter $\nu$, and $n=1$, model (\ref{eq: modeling framework}) resembles to \citet{gromenko2012} and \citet{gromenko2013nonparametric} for spatially indexed functional data. (iii) If $\xi_{ik}(T_{ij}) = \sum_{l \geq 1} \zeta_{ikl} \psi_{ikl}(T_{ij})$ with eigenfunctions $\psi_{ikl}(T)$'s and the corresponding coefficients $\zeta_{ikl}$'s, then model (\ref{eq: modeling framework}) is similar to CM when $\phi_{k}(\cdot|T) = \phi_{k}(\cdot)$ for all $k$ and $t$.

The model proposed in \citet{chen2013repeated} is, in fact, closely related to model (\ref{eq: modeling framework}), but with one fundamental difference. Model (\ref{eq: modeling framework}) uses a time-invariant basis function, $\{\phi_k(\cdot)\}_k$, while \citet{chen2013repeated} use a time-varying basis function, $\{\phi_k(\cdot|T_{ij})\}_k$. This key difference actually leads to the major advantages of the modeling framework proposed in this paper. First, by using a time-invariant basis function, the basis coefficients, $\xi_{ik}(T_{ij})$, extract the low dimensional features of these massive data with complex dependence structure. The longitudinal dynamics is emphasized only through the time-varying coefficients $\xi_{ik}(T_{ij})$ of (\ref{eq: modeling framework}), and thus this perspective makes the study of the process dynamics easier to understand. Second, the estimation approach of CM requires to obtain eigenbasis functions $\{\phi_k(\cdot|T_{ij})\}_k$ at each $T_{ij}$, by employing functional principal component analysis (FPCA) multiple times; our method requires to obtain only one set of basis functions, $\{\phi_k(\cdot)\}_k$. As the set $\{T_{ij}: i,j\}$ is dense in $\mathcal{T}$, their estimation method involves a rather complex implementation and is computationally burdensome. In contrast, our method has computational advantages; the estimation is based on two dimensional smoother, which is faster than the three dimensional smoother required by the methods of \citet{chen2013repeated}. 

Using time-invariant basis functions has many appealing advantages. However, selecting a time-invariant basis is nontrivial. One option is to use a pre-specified basis but it brings along the challenging issues of deciding which basis to use, as well as of selecting the optimal number of basis functions. Data-driven basis is another option though there is no obvious way to select it. Here we propose to determine $\{\phi_k(\cdot)\}_k$ using an appropriate \textit{marginal} covariance operator $\Sigma(s,s') = \int_{\mathcal{T}} c((s,T),(s',T)) g(T) dT$, where $c((s,T), (s',T'))$ is the covariance function of the process, $X(s, T)$, and $g(T)$ is the sampling density of $T$'s. In Section \ref{sec: model} we show that the proposed basis, $\{\phi_k(\cdot)\}_k$, has optimal properties with respect to some appropriately defined criterion. From this view point, the model representation (\ref{eq: modeling framework}) is optimal. The idea of using the eigenbasis of the pooled covariance has been discussed recently in \citet{jiang2010covariate} and \citet{pomann2013two} for the case of independent functional data. 

The rest of paper is organized as follows. Section \ref{sec: model} introduces the proposed modeling framework. Section \ref{sec: estimation} describes the estimation methods and implementation. The methodology is studied theoretically in Section \ref{sec: theoretical properties} and then numerically in Section \ref{sec: simulation}. Section \ref{sec: application} discusses the application to the tractography DTI data. The paper concludes with a brief discussion in Section \ref{sec: discussion}.


\section{Modeling longitudinal functional data} \label{sec: model}

Let $[\{T_{ij}, Y_{ij}(s_r): r=1,\ldots, R \}: j=1,\ldots,m_i, ]$ be the data for the $i$th subject, where $Y_{ij}(\cdot)$ is the $j$th profile recorded at visit time, $T_{ij}$ for subject $i$, and each profile is observed at the fine grid of points $\{s_1,\ldots, s_R\}$. For notation simplicity we use the generic index $s$ instead of $s_r$ and refer to $s$ by `functional argument'. We assume that, for each subject $i$, the number of observed curves, $m_i$, is relatively small, and the visit times, $\{T_{ij}\}_j$ are random in a closed and compact interval, $\mathcal{T}$. It is assumed that the set of time points of all subjects, $\{T_{ij}: \text{ for all } i, j\}$, is dense in $\mathcal{T}$; we call the generic time $T$  by `longitudinal argument'. Without loss of generality, we set $\mathcal{S}=\mathcal{T}=[0,1]$. 

Now consider model (\ref{eq: modeling framework}) 
where our primary goal is to model the complex covariance function of the random process $X_i(s, T_{ij})=X_{ij}(s)$; we use this notation interchangeably. In our exposition it is helpful to represent the noise process $\epsilon_{ij}(\cdot)$ as the sum of two well identified components: one square integrable smooth component $\epsilon_{1,ij}(s)$, which has a covariance function $\Gamma(s,s')=\rm{cov}\{\epsilon_{1,ij}(s), \epsilon_{1,ij}(s')\}$ that varies smoothly over $s,s'\in [0,1]$ and a white noise component $\epsilon_{2,ij}(s)$ with covariance $\rm{cov}\{\epsilon_{2,ij}(s), \epsilon_{2,ij}(s')\} = \sigma^2 $ if $s=s'$ and $0$ otherwise; only the component $\epsilon_{1,ij}(s)$ is relevant for the nontrivial process dependence.

Let $c((s,T),(s',T')) = \text{E}[X_{i}(s,T)X_{i}(s',T')]$ be the covariance function of the process $X_i(\cdot,\cdot)$ and $g(\cdot)$ be the sampling density of $T$. Define the \textit{marginal} covariance function induced by $X_i$'s by $\Sigma(s,s') = \int c((s,T), (s',T))g(T) dT$; in Section \ref{sec: theoretical properties} we show that $\Sigma(s,s')$ is a \textit{proper} covariance function \citep{horvath2012inference}. Denote by $W_i(s,T_{ij}) = X_i(s, T_{ij}) + \epsilon_{1,ij}(s)$; $W_i$ is a bivariate random process defined on $[0,1]\times [0,1]$ and its induced marginal covariance is $\Xi(s,s') =\Sigma(s,s') + \Gamma(s,s') $. Let $\{ \phi_k(s), \lambda_k\}_{k}$ be the pair of eigenfunctions and eigenvalues from the spectral decomposition of $\Xi(s,s') $, where $\{ \phi_k(\cdot): k\}$ forms an orthogonal basis in $L^2[0,1]$ and $\lambda_1 \geq  \lambda_2 \geq \ldots \geq 0$. Using arguments similar to the standard FPCA, the eigenbasis functions $\{\phi_k(\cdot): k=1, 
\ldots,K\}$ are optimal in the sense that they minimize the following weighted mean square error: $\text{MSE}(\theta_{1}(\cdot), \ldots, \theta_{K}(\cdot)) = \int_{0}^1\text{E}\| W_{i}(\cdot, T) - \sum_{k=1}^{K}\theta_k(\cdot) <W_{i}(\cdot, T), \theta_{k}(\cdot)> \|^2 g(T) dT,$ where $<f_{1}(\cdot), f_{2}(\cdot)> = \int_0^1 f_{1}(s)f_{2}(s)ds$ is the usual inner product in $L^2[0,1]$.

Using the orthogonal basis in $L^2[0,1]$ $\{\phi_k(\cdot)\}_k$, we can represent the square integrable smooth process $W_i(\cdot, T)$ as
$
W_{i}(s, T_{ij}) = \sum_{k=1}^{\infty} \xi_{W,ijk} \phi_k(s),
$
where 
\begin{equation} \label{eq: first fpca1}
\xi_{W,ijk} = \int W_{i}(s, T_{ij})\phi_k(s) ds =  \xi_{ik}(T_{ij})+ e_{ijk},
\end{equation}
and $\xi_{W,ijk}$ are not necessarily uncorrelated over $k$. Here $\xi_{ik}(T_{ij}) =\int X_{i}(s, T_{ij})\phi_k(s) ds $ and $ e_{ijk}= \int \epsilon_{1,ij}(s)\phi_k(s)ds$ are specified by the definition of $W_i$; for fixed $k$ these terms are mutually independent due to the independence of the processes $X_i$ and $\epsilon_{1,ij}$. For each $k$, one can easily show that, $\xi_{ik}(\cdot)$ is a smooth zero-mean random process in $L^2[0,1]$ and is iid over $i$; by an abuse of notation in what follows we use interchangeably $\xi_{ik}(T_{ij})=\xi_{ijk}$. Furthermore $ e_{ijk}$ are zero-mean iid random variables over $i,j$; denote by $\sigma^2_{e,k} $ their finite variance. The above representation allows us to recover the latent signal $X_i(\cdot, T_{ij})$ and error $\epsilon_{1,ij}(\cdot)$ as $X_{i}(s, T_{ij}) = \sum_{k=1}^{\infty} \xi_{ik}(T_{ij}) \phi_k(s) \text{ and } \epsilon_{1,ij}(s) = \sum_{k=1}^{\infty} e_{ijk} \phi_k(s)$

The main difference between the proposed work and \citet{chen2013repeated} is the use of the marginal covariance function, $\Sigma(s,s')$; CM use the conditional covariance function $\Sigma(s,s'|T)$. The proposed approach reduces computational burden substantially, by avoiding the three dimensional covariance function and the spectral decomposition of $\Sigma(s,s'|T)$ at every $T$. Also using time-invariant basis functions is critical in reducing the dimensionality of the data in a way that captures the dependence over visit time; the longitudinal dependence (over $T$) can be studied through $\xi_{ijk}=\xi_{ik}(T_{ij})$'s.

One way to model the dependence of the basis coefficients, $\xi_{ijk}$, is by using common techniques in longitudinal data analysis; for example by assuming a parametric covariance structure. As we discussed in Section \ref{sec: introduction}, this leads to models similar to \citet{greven2010, gromenko2012, gromenko2013nonparametric}. We consider this approach in the analysis of the DTI data, Section \ref{sec: application}. Another approach is to assume a nonparametric covariance structure and employ a common functional data analysis technique. We detail the latter approach in this section. 

Specifically, for each $k\geq 1$ denote by $G_{k}(T,T') = \textrm{cov}\{\xi_{ik}(T), \xi_{ik}(T')\}$ the smooth covariance function in $[0,1]\times [0,1]$. Mercer's theorem 
provides the following convenient spectral decomposition $G_{k}(T,T') = \sum_{l\geq 1} \eta_{kl}\psi_{kl}(T) \psi_{kl}(T')$, where $\eta_{k1} \geq \eta_{k2} \geq  \ldots \geq 0  $ and $\{ \psi_{kl}(\cdot)\}_{l\geq 1}$ is an orthogonal basis in $L^2[0,1]$. Using the Karhunen-Lo\`{e}ve (KL) expansion, we represent $\xi_{ik}(\cdot)$ as:
\begin{equation} \label{eq: second fpca}
\xi_{ik}(T_{ij}) = \sum_{l=1}^{\infty} \zeta_{ikl} \psi_{kl}(T_{ij}), \qquad  \zeta_{ikl} = \int \xi_{ik}(T) \psi_{kl}(T) dT,
\end{equation}
where $\zeta_{ikl}$ have zero-mean, variance equal to $\eta_{kl}$, and are uncorrelated over $l$.

By collecting all the components, we represent the model (\ref{eq: modeling framework}) as $Y_{ij}(s) = \mu(s, T_{ij}) + \sum_{k=1}^{\infty} \sum_{l=1}^{\infty} \zeta_{ikl} \psi_{kl}(T_{ij})\phi_k(s) + \epsilon_{ij}(s)$, for $\epsilon_{ij}(s)=\sum_{k=1}^{\infty} e_{ijk}\phi_k(s) + \epsilon_{2,ij}(s)$. In practice we truncate this expansion. Let $K$ and $L_k$'s such that $Y_{ij}(s)$ is well approximated by the following truncated model based on the leading $K$ and $\sum_k L_k$ respective basis functions 
\begin{equation}\label{eq: truncated final model} 
Y_{i}(s, T_{ij})  = \mu(s, T_{ij}) + \sum_{k=1}^{K} \sum_{l=1}^{L_k} \zeta_{ikl} \psi_{kl}(T_{ij})\phi_k(s) + \epsilon_{ij}(s),
\end{equation}
where $\epsilon_{ij}(s) \approxeq \sum_{k=1}^{K} e_{ijk}\phi_k(s) + \epsilon_{2,ij}(s)$.
The truncated model (\ref{eq: truncated final model}) gives a parsimonious representation of the complex dependent longitudinal functional data; it allows the study of its dependence through 
two sets of eigenfunctions: one dependent only on $s$ and one only on visit times, $T_{ij}$. This approach involves two main challenges: first, determining consistent estimators of the orthogonal basis functions $\phi_k(\cdot)$'s, and second estimating the covariance functions $G_k(\cdot, \cdot)$ of the time-varying coefficients $\xi_{ik}(T)$, when the quantities $W_i(s, T_{ij})$'s are not directly observable.

\section{Estimation of model components} \label{sec: estimation}
Next we discuss estimation of all model components: Section \ref{subsec: estimation (mean)} describes the estimation of the mean function. Section \ref{subsec: estimation (1st FPCA)} presents the estimation of the marginal covariance function and of the eigenfunctions $\phi_k(\cdot)$'s, and Section \ref{subsec: estimation (2nd FPCA)} presents the estimation of the time-varying basis coefficients $\xi_{ik}(\cdot)$'s. Prediction of $Y_{i}(s, T) $ is detailed in Section \ref{subsec: estimation (predicted)}.  

\subsection{Step 1: Mean function $\mu(s,T)$} \label{subsec: estimation (mean)}

In this paper we estimate the mean function, $\mu(s,T)$, using bivariate smoothing via bivariate tensor product splines \citep{wood2006a} of the pooled data $\{Y_{ijr} = Y_{ij}(s_r): i = 1,\ldots,n, j=1,\dots,m_i \text{ and } r= 1,\ldots,R\}$. Consider two univariate B-spline bases, $\{B_{s,1}(s),$ $\ldots, B_{s,d_s}(s)\}$ and $\{B_{T,1}(T),\ldots, B_{T,d_T}(T)\}$, defined on $\mathcal{S}=[0,1]$ and $\mathcal{T}=[0,1]$, respectively, where $d_s$ and $d_T$ are their respective dimensions. The mean surface is represented as a linear combination of a tensor product of the two univariate B-spline bases $\mu(s,T) = \sum_{q_1=1}^{d_s} \sum_{q_2=1}^{d_T}$ $ B_{s,q_1}(s)B_{T,q_2}(T)\beta_{q_1q_2} = \mathbf{B}(s,T)^T\boldsymbol{\beta}$, where $\mathbf{B}(s,T)$ is the known $d_sd_T$-dimensional vector of $B_{s,q_1}(s)B_{T,q_2}(T)$'s, and $\boldsymbol{\beta}$ is the vector of unknown parameters, $\beta_{q_1q_2}$'s. The bases dimensions, $d_s$ and $d_T$, are set to be sufficiently large to accommodate the complexity of the true mean function, and roughness of the function is controlled through the size of the curvature in each direction separately, i.e. $\iint \{\partial^2\mu(s,T)/\partial s^2\}^2 dTds = \boldsymbol{\beta}^T(\mathbf{P_s}\otimes\mathbf{I_{d_T}})\boldsymbol{\beta}$ in direction $s$, and $\iint \{\partial^2\mu(s,T)/\partial T^2\}^2 dTds = \boldsymbol{\beta}^T(\mathbf{I_{d_s}} \otimes \mathbf{P_T})\boldsymbol{\beta}$ in $T$. The penalized criterion to be minimized is

\begin{equation}
 \sum_{i,j,r}\big[ Y_{ijr} - \mathbf{B}(s_r,T_{ij})^T\boldsymbol{\beta}   \big] ^2 + \boldsymbol{\beta}^T(\lambda_s\mathbf{P_s}\otimes\mathbf{I_{d_T}} + \lambda_T \mathbf{I_{d_s}} \otimes \mathbf{P_T})\boldsymbol{\beta},
\end{equation}
where $\lambda_s$ and $\lambda_T$ are smoothing parameters that control the trade-off between the smoothness of the fit and the goodness of fit. The smoothing parameters can be selected by the restricted maximum likelihood (REML) or generalized cross-validation (GCV). It follows that the estimated mean function $\widehat{\mu}(s,T) = \mathbf{B}(s,T)^T\boldsymbol{\widehat{\beta}}$. 

While this method is a very popular smoothing technique of bivariate data, other available bivariate smoothers can be used to estimate the mean $\mu(s,T)$; for example, kernel-based local linear smoother \citep{hastie2009elements}, bivariate penalized spline smoother \citep{marx2005multidimensional} and the sandwich smoother \citep{xiao2013fast}. The sandwich smoother \citep{xiao2013fast} is especially useful in the case of very high dimensional data for its appealing computational efficiency, in addition to its estimation accuracy.

\subsection{Step 2: Marginal covariance. Data-driven orthogonal basis} \label{subsec: estimation (1st FPCA)}

Once the mean function is estimated, let $\widetilde{Y}_{ijr} = Y_{ijr} - \widehat{\mu}(s_r,T_{ij})$ be the demeaned data. We use the demeaned data to estimate the marginal covariance function induced by $W_{i}(s, T_{ij})$, $\Xi(s,s') = \Sigma(s,s') + \Gamma(s,s')$. The estimation of $\Xi(s,s')$ consists of two steps. In the first step, a raw covariance estimator $\widetilde{\Xi}(s,s')$ is obtained; the pooled sample covariance is a suitable choice, if all the curves are observed on the same grid of points: 
\begin{equation} \label{eq: raw marginal covariance}
\widetilde{\Xi}(s_r,s_{r'}) = \sum_{i=1}^{n} \sum_{j=1}^{m_i} \widetilde{Y}_{ijr} \widetilde{Y}_{ijr'}/\big(\sum_{i=1}^{n} m_i\big).
\end{equation}
Because we assume that each observation, $Y_{ijr}$, is observed with white noise, $\epsilon_{2,ij}(s_r)$, the `diagonal' elements of the sample covariance, $\widetilde{\Xi}(s_r,s_{r})$, are inflated by the variance of the noise, $\sigma^2$. In the second step, the preliminary covariance estimator is smoothed by ignoring the `diagonal' terms; see also \citet{staniswalis1998nonparametric} and \citet{yao2005functional}) who used similar technique for the case of independent functional data. In our simulation and data application we use the sandwich smoother \citep{xiao2013fast}. To ensure the positive semi-definiteness of the estimator any negative eigenvalues are zero-ed. The resulting smoothed covariance function, $\widehat{\Xi}(s,s')$, is used as an estimator of $\Xi(s,s')$. In Section \ref{sec: theoretical properties}, we show that $\widehat{\Xi}(s,s')$ is an unbiased and consistent estimator of $\Xi(s,s')$ in two settings: 1) the data are observed fully and without measurement errors, i.e. $\epsilon_{ij}(s) \equiv 0 $ and 2) the data are observed fully and with measurement error of type $\epsilon_{1,ij} (s)$, i.e $\epsilon_{ij}(s) \equiv \epsilon_{1,ij}(s)$. 


Let $\{\widehat{\phi}_k(s), \widehat{\lambda}_k\}_k $ be the pairs of eigenvalues/eigenfunctions obtained from the spectral decomposition of the estimated covariance function, $\widehat{\Xi}(s,s')$. The truncation value $K$ is determined based on pre-specified percentage of variance explained (PVE); specifically, $K$ can be chosen as the smallest integer such that $ \big(\sum_{k=1}^{K} \widehat{\lambda}_k/\sum_{k=1}^{\infty} \widehat{\lambda}_k\big)$ is greater than the pre-specified PVE \citep{di2009multilevel,staicu2010fast}.

\subsection{Step 3: Covariance modeling of the longitudinal varying components} \label{subsec: estimation (2nd FPCA)}

Let $\widetilde \xi_{W,ijk} = \int \widetilde Y_{ij}(s) \widehat \phi_k(s) ds $ be the projection of the $j$th repeated demeaned curve of the $i$th subject onto the direction $\widehat \phi_k(\cdot)$ for $k=1, \ldots, K$. Since $\widetilde Y_{ij}(s)$, is observed at dense grids of points $\{s_r: r=1,\ldots,R\}$ in $\mathcal{S}$ for all $i$ and $j$, $\widetilde \xi_{W,ijk}$ will be approximated accurately through numerical integration. It is easy to see that the version of  $\widetilde \xi_{W,ijk}$ that uses $ \mu(s, T_{ij})$ in place of $ \widehat \mu(s, T_{ij})$ and $\phi_k(s)$ in place of $\widehat \phi_k(s)$ converges to $\xi_{W,ijk}$ with probability one, as $R$ diverges. The time-varying coefficients $\widetilde \xi_{W,ijk} $ are proxy measurements of $ \xi_{ijk} = \xi_{ik}(T_{ij}) $ and will be used to study the process dynamics. Estimation of their covariance is important as it describes the process temporal dependence but also because it allows prediction at unobserved time points $T\in [0,1]$. One alternative is to stack the coefficients $\widetilde \xi_{W,ijk} $ over $k$ and study the time-dependence of the resulting $K$-dimensional vector. We take a simpler way and study the temporal dynamics separately for each $k$.


Let $\{(T_{ij}, \widetilde \xi_{W,ijk}): j=1, \ldots, m_i\}_i$ be the `observed data' that will be used to estimate the temporal covariance. 
A possible approach is to assume a parametric model for the latent temporal covariance, such as AR(1) or a random effects model framework; such approach would be preferable when there are very few curve measurements per subject and the longitudinal design is balanced. We discuss random effects model for estimating the longitudinal covariance in the data application. Here we consider a more flexible approach and estimate the covariance nonparametrically. Specifically, recall that $G_k(T, T')=\textrm{cov}\{\xi_{ik}(T), \xi_{ik}(T')    \} $ and our framework as described in (\ref{eq: first fpca1}) is common for modeling discretely observed functional data: $\xi_{W,ijk} = \sum_{l=1}^\infty \zeta_{ikl} \psi_{kl}(T_{ij})  + e_{ijk}$ where 
$\zeta_{ikl}$ are random variables with zero mean and $\eta_{kl}$ variances, 
$ \{\psi_{kl}(T), \eta_{kl}\}_l$ are the pairs of eigenfunctions and eigenvalues of the covariance $G_k$, and 
$e_{ijk}$'s are iid with zero-mean and variance equal to $\sigma^2_{e,k}$. Though we do not observe $\xi_{W,ijk}$'s we use the proxy $\widetilde \xi_{W,ijk}$'s and estimate $G_k$ by employing FPCA techniques for sparse functional data \citep{yao2005functional}. 

For completeness we summarize the approach below. Following \citet{yao2005functional}, we first obtain the raw sample covariance, $\widetilde{G}_{ik}(T_{ij}, T_{ij'}) =  \widetilde \xi_{W,ijk}  \widetilde \xi_{W,ijk} $. Then the estimated smooth covariance surface, $\widehat{G}_{k}(T, T')$, is obtained by using bivariate smoothing of $\{ (T_{ij}, T_{ij'}), \widetilde{G}_{ik}(T_{ij}, T_{ij'}) :i, j\neq j'\}$. Kernel-based local linear smoothing \citep{yao2005functional} or penalized tensor product spline smoothing \citep{wood2006a} can be used at this step. The diagonal terms $\{\widetilde{G}_{ik}(T_{ij}, T_{ij'}): i, j=j'\}$ are removed because the noise $e_{ijk}$ present in the proxy $\xi_{W,ijk}$ leads to inflated variance function. Let $\{\widehat{\psi}_{kl}(\cdot), \widehat{\eta}_{kl}\}_l$ be the pairs of eigenvalues/eigenfunctions of the estimated covariance surface, $\widehat{G}_{k}(T,T')$. The truncation value, $L_k$, is determined based on pre-specified PVE; using similar ideas as in Section \ref{subsec: estimation (1st FPCA)}. The variance $\sigma^2_{e,k}$ is estimated from the difference between the variance along the diagonal, or a smooth estimator of the raw diagonal terms, and the predicted covariance for $T=T'$, $\widehat G_k(T, T)$; \citet{yao2005functional} discusses an alternative that dismisses the terms at the boundary when estimating the error variance. Once the basis functions $\{\psi_{kl}(\cdot)\}_{l=1}^{L_k}$, eigenvalues $\eta_{kl}$'s, and error variance $\sigma^2_{e,k}$ are estimated, the above model framework can be viewed as a mixed effects model and the random components $\zeta_{ikl}$ can be predicted using conditional expectation and a jointly Gaussian assumption for $\zeta_{ijk}$'s and $e_{ijk}$'s. In particular, 
$\widehat{\zeta}_{ikl} = \widehat {\text{E}[\zeta_{ikl}|\widetilde {\boldsymbol{\xi}}_{W,ik} ]} = 
\widehat \eta_{kl} \widehat {\boldsymbol{\psi}}_{ikl}^T\widehat {\boldsymbol{\Sigma}}_{\mathbf{\xi}_{ik}} \ \widetilde {\boldsymbol{\xi}}_{W,ik} 
 $, where
$\widehat {\boldsymbol{ \psi}}_{kl} = \{ \widehat  {\psi}_{kl}(T_{i1}), \ldots,  \widehat {\psi}_{kl}(T_{im_i}) \}^T$ is the $m_i$-dimensional column vector of the evaluations of $\widehat {\psi}_{kl}(\cdot)$ at $\{T_{ij}:j=1,\ldots, m_i\}$, $\widehat {\boldsymbol{\Sigma}}_{\mathbf{\xi}_{ik}}$ is a $m_i\times m_i$ -  matrix with $(j, j')$th element equal to $\widehat G_k(T_{ij}, T_{ij'})+\widehat \sigma_{e,k}^2\delta_{jj'}$, for $\delta_{jj'}=1$ of $j=j'$ and $0$ otherwise, and $\widetilde {\boldsymbol{\xi}}_{W,ik}$ is the $m_i$ dimensional column vector of $\widetilde \xi_{W, ijk}$'s. The predicted time-varying coefficients corresponding to a generic time $T$ are obtained as 
\begin{equation} \label{eq: estimation of score trajectories}
\widehat{\xi}_{ik}(T) = \sum_{l=1}^{L_k} \widehat{\zeta}_{ikl} \widehat{\psi}_{kl}(T).
\end{equation}

\citet{yao2005functional} proved the consistency of the estimators of the model components and the predicted score trajectories when $\xi_{W,ijk}$'s are observed directly. In Section \ref{sec: theoretical properties} we extend these results to the case when the proxy $\widetilde  \xi_{W,ijk}$'s are used instead, and when the observed curves $Y_{ij}(\cdot)$ are fully observed and the measurement error is of the type $\epsilon_{ij}(s) = \epsilon_{1,ij}(s)$; i.e. the data are observed with smooth error. 

\subsection{Step 4: Trajectories reconstruction} \label{subsec: estimation (predicted)}

We are now able to predict the full response curve at any time point $T \ in[0,1]$ by: 
\begin{equation}  \label{eq: predicted trajectory}
\widehat{Y}_i(s, T) = \widehat{\mu}(s,T) + \sum_{k=1}^{K} \widehat{\xi}_{ik}(T) \widehat{\phi}_k(s), \qquad s\in[0,1]; 
\end{equation}
where $\widehat{\xi}_{ik}(T) = \sum_{l=1}^{L_k} \widehat{\zeta}_{ikl} \widehat{\psi}_{kl}(T)$. 

\subsection{Implementation using available softwares} \label{subsec: estimation (software)}

An important advantage of the proposed approach is that its implementation can be carried using 
available software.

\begin{enumerate}[label= \textbf{Step \arabic*.}]
\item Estimate the smooth mean function $\widehat \mu(s, T)$ using the sandwich smoother \citep{xiao2013fast} (the \verb|fbps| function in \verb|R| \citep{Rsoftware} package \verb|refund| \citep{Rrefundpackage}) or using the penalized tensor product spline smoothing (the \verb|gam| and \verb|te| functions in \verb|R| \citep{Rsoftware} package \verb|mgcv| \citep{wood2011faststable}).

\item Estimate the smooth covariance function $\widehat \Xi(s,s')$ with the demeaned data (dense design) using the sandwich smoother \citep{xiao2013fast} and get the eigenfunctions $\widehat \phi_k(s)$ using the \verb|fpca.face| function in the \verb|refund| package \citep{Rrefundpackage}. The default option of this function also provides $\widetilde \xi_{W, ijk}$'s.

\item For each $k$, carry out FPCA of $\{ T_{ij}, \widetilde \xi_{W, ijk}:i, j\}$. There are several available options for implementation: \verb|fpca.sc| function in the \verb|refund| package \citep{Rrefundpackage} and \verb|fpca.mle| and \verb|fpca.pred| functions in the \verb|FPCA| package \citep{peng2009geometric, james2000principal} in {\tt{R}}. Alternatively one can use the \verb|FPCA| function \citep{yao2005functional} in the \verb|MATLAB| \citep{MATLAB:2014} package \verb|PACE| \citep{yao2005functional} available at \verb|http://www.|\verb|stat.ucdavis.edu/PACE/|.

\item Determine the predicted trajectories using equation (\ref{eq: predicted trajectory}).
\end{enumerate}


\section{Theoretical properties} \label{sec: theoretical properties}

We discuss now the asymptotic properties of the model components estimators and the predicted trajectories. Our study is based on the assumption of a sparse longitudinal design (modest number of repeated curve measurements per subject) and dense functional design; this requires new techniques than the ones commonly used for theoretical investigation of repeated functional data (such as \cite{chen2013repeated}).

Throughout this section we assume that response trajectories, $Y_{ij}(\cdot)$'s, have zero-mean and are observed fully as a function over the domain, $\mathcal{S} = [0,1]$. Section \ref{subsec: theoretical results with no errors} discusses the main theoretical results when data, $Y_{ij}(s)$'s, are observed without error, i.e. $\epsilon_{ij}(s)\equiv 0$ for $s\in[0,1] $. Section \ref{subsec: theoretical results with error process} extends the findings to the case when the observed data are corrupted with a smooth error process $\epsilon_{ij}(s)\equiv \epsilon_{1,ij}(s)$; i.e. there is no white noise measurement error. The proofs of the results are sketched in the Appendix and described to greater detail in the Supplementary Material. Section \ref{subsec: theoretical results discussion} discusses on how to relax some of the  assumptions. 
For clarity, throughout this section we use $\mathcal{S}$ and $\mathcal{T}$ to distinguish between the domains for the functional argument and longitudinal one, respectively. 

We assume that the bivariate process $X_{i}(s,T_{ij})$ is a realization of a true random process, $X(s,T)$, with zero-mean and smooth covariance function, $ c((s,T), (s',T')) = \textrm{cov}\{X(s,T), X(s',T')\}$, which furthermore satisfies some regularity conditions. 
\begin{enumerate} [label= (A\arabic*.) ]

	\item $X = \{ X(s,T): (s,T) \in \mathcal{S} \times \mathcal{T} \}$ is a square integrable element of the $L^2( \mathcal{S}\times  \mathcal{T})$, 
	i.e. $\text{E}[\iint X^2(s,T)dsdT] < \infty$, where $\mathcal{S}$ and $\mathcal{T}$ are compact sets. \label{assmp: square int X}
	
	\item The sampling density $g(T)$ is continuous and $\sup_{T\in \mathcal{T}} |g(T)|<\infty$.\label{assmp: reg con g(T)}

\end{enumerate}

We first verify that the marginal covariance, $\Sigma(s,s')$, is a \textit{proper} covariance function \citep[p.24]{horvath2012inference}. 
Under the assumptions \ref{assmp: square int X} and \ref{assmp: reg con g(T)}, the marginal covariance function, $\Sigma(s,s')= \int c\{(s,T),(s',T)\} g(T) dT$, (i) is symmetric, (ii) is positive definite, and (iii) have eigenvalues satisfying that $\sum_{k=1}^{\infty} \lambda_{k}$ is finite.

\subsection{Response curves measured without error} \label{subsec: theoretical results with no errors}

Assume that there is no error $ \epsilon_{ij}(s)\equiv 0$ and thus $Y_{ij}(s) = X_{i}(s, T_{ij})$ for $s\in \mathcal{S}$. Then the sample covariance of $Y_{ij}(s)$ is $\widehat{\Sigma}(s,s')=\sum_{i=1}^{n} \sum_{j=1}^{m_i} Y_{ij}(s)Y_{ij}(s')/$ $(\sum_{i=1}^{n} m_i)$. We first show that $\widehat{\Sigma}(s,s')$ is an unbiased estimator of the true marginal covariance, $\Sigma(s,s)$, and then we prove that it is a consistent estimator. The following assumptions are important in our theoretical development.

\begin{enumerate} [resume, label= (A\arabic*.) ]
	\item $\text{E}[X(s,T)X(s',T)X(s,T')X(s',T')] < \infty$ for arbitrary $s,s'\in \mathcal{S}$ and $T,T'\in \mathcal{T}$  . \label{assmp: E[A(T)A(T')|T,T']} 
	\item $\text{E}\big[ \|X(\cdot, T) \|^4\big] < \infty $ for each $T\in \mathcal{T}$. \label{assmp: 4th moment Y}
\end{enumerate}

Conditions \ref{assmp: E[A(T)A(T')|T,T']} and \ref{assmp: 4th moment Y} regard the moment behavior of the latent process $X$ and are commonly used in functional data analysis \citep{yao2005functional,chen2013repeated}. Theorem \ref{th: consistency of Sigma} gives the asymptotic properties of the marginal covariance estimator $\widehat \Sigma(s,s')$.

\begin{thm}\label{th: consistency of Sigma}
	Assume \ref{assmp: square int X} - \ref{assmp: E[A(T)A(T')|T,T']} hold. Then we have that for each $(s,s')$, $
	|\widehat{\Sigma}(s,s') - \Sigma(s,s')| \xrightarrow{\text{  p  }} 0 $
	as $n$ diverges. If the additional assumption 
	\ref{assmp: 4th moment Y} holds, then we have that 
	\begin{equation} \label{eq: consistency of HSnorm}
	\| \widehat{\Sigma}(\cdot, \cdot) - \Sigma(\cdot, \cdot) \|_{s} \xrightarrow{\text{  p  }} 0 \text{ as } n \longrightarrow \infty,
	\end{equation} 
	where $\| k(\cdot, \cdot)\|_{s} = \{ \iint k^2(s,s') dsds' \}^{1/2}$ is the Hilbert-Schmidt norm of $k(\cdot,\cdot)$.  
\end{thm}

By the consistency result (\ref{eq: consistency of HSnorm}) and Theorem 4.4 and Lemma 4.3 of \citet[p.104]{bosq2000linear}, it follows that the eigen-elements of $\widehat{\Sigma}(s,s')$ are consistent estimators of the corresponding eigen-elements of $\Sigma(s,s')$, if the eigenvalues are neither crossing nor the same.

\begin{enumerate} [resume, label= (A\arabic*.) ]
	\item Let $a_k = (\lambda_1 - \lambda_2)$ for $k=1$ and $a_k =\text{max}[(\lambda_{k-1}- \lambda_{k}), (\lambda_{k}-\lambda_{k+1}))]$ otherwise, where $\lambda_k$ is the $k$th largest eigenvalues of $\Sigma(s,s')$. Assume that $0<a_k<\infty$ and $\lambda_k > 0$ for all $k$ (No crossing or ties among eigenvalues). \label{assmp: no crossing marginal eigenvalues}

\end{enumerate}

\begin{cor} \label{cor: consistency of eigencomponents of mFPCA}
	Under the assumptions \ref{assmp: square int X},\ref{assmp: reg con g(T)}, \ref{assmp: 4th moment Y}, and \ref{assmp: no crossing marginal eigenvalues}, for each $k$
	\begin{equation} \label{eq: consistency of eigencomponents of mFPCA}
	|\widehat{\lambda}_k - \lambda_k| \xrightarrow{\text{  p  }} 0, \text{ and }\|\widehat{\phi}_k(\cdot)-\phi_k(\cdot)\|_s \xrightarrow{\text{  p  }} 0 \qquad \text{as $n$ diverges.}
	\end{equation}
	
\end{cor}

Next, we focus on the estimation of the covariance that describes the longitudinal dynamics. If $\xi_{W,ijk}$'s were available to estimate the covariance function, $G_{k}(T_{ij}, T_{ij'})= \textrm{cov}(\xi_{ik}(T_{ij}), \xi_{ik}(T_{ij'}))$, then the consistency of the model components follows trivially from the FPCA properties developed in \citet{yao2005functional}. However, $\xi_{W,ijk}$'s are not directly observed; instead available are the proxy time-varying coefficients $\widetilde{\xi}_{W,ijk}= \int Y_{i}(s,T_{ij})\widehat{\phi}(s) ds$. We first show the uniform consistency of $\widetilde{\xi}_{W,ijk}$, then use this result to show that the estimator of $G_{k}(T_{ij}, T_{ij'})$ based on $\widetilde{\xi}_{W,ijk}$ is asymptotically identical to that based on $\xi_{W,ijk}$. Consistency results of the remaining model components follow directly from \citet{yao2005functional}. The uniform consistency of $\widetilde{\xi}_{ijk}$ requires $\text{sup}_{j,s} |Y_i(s,T_{ij})|$ to be bounded almost surely, and this is ensured by assuming \ref{assmp: Y asymptotically bounded}. The Gaussian assumption \ref{assmp: jointly gaussian} is needed for the consistency of $\widehat{\zeta}_{ikl}$ obtained using the PACE method of \citet{yao2005functional}.

\begin{enumerate} [resume, label= (A\arabic*.) ]
\item $\text{E}\big[ \text{sup}_{s,T} |X(s,T)|^a \big] \leq M^a$ for a constant, $M > 0$, and an arbitrary integer, $a \geq 1$; This is equivalent to assume that $X(s,T)$ is absolutely bounded almost surely. \label{assmp: Y asymptotically bounded}
	\item Let $b_{kl} = (\eta_{k1} - \eta_{k2})$ for $l=1$ and $b_{kl} =\text{max}[(\eta_{k(l-1)}- \eta_{kl}), (\eta_{kl}-\eta_{k(l+1)}))]$ otherwise, where $\eta_{kl}$ is the $l$th largest eigenvalues of $G_k(T,T')$. Assume that $0<b_{kl}<\infty$ and $\eta_{kl} > 0$ for all $k$ and $l$ (No crossing or ties among eigenvalues). \label{assmp: no crossing second eigenvalues}
	
	\item $\zeta_{ikl}$ and $e_{ijk}$ are jointly Gaussian. \label{assmp: jointly gaussian}
\end{enumerate}

\begin{thm} \label{th: consistency of xi and G}
	Under the assumptions \ref{assmp: square int X}, \ref{assmp: reg con g(T)}, \ref{assmp: 4th moment Y} - \ref{assmp: Y asymptotically bounded}, for each $k$
	\begin{equation} \label{eq: consistency of xi}
	\text{sup}_{j} | \widetilde{\xi}_{W,ijk} - \xi_{W,ijk}| \xrightarrow{\text{  p  }} 0
	\end{equation}
	and
	\begin{equation} \label{eq: HSnorm consistency of G}
	\| \widehat{G}_{k}(\cdot,\cdot) - G_{k}(\cdot,\cdot)  \|_s \xrightarrow{\text{  p  }} 0
	\end{equation}
	as $n$ diverges. In fact a stronger result also holds, namely the uniform convergence of $\widehat{G}_{k}(T,T')$. Specifically, $\text{sup}_{T} |\widehat{G}_{k}(T,T') - G_{k}(T,T')| \xrightarrow{\text{  p  }} 0$ as $n$ diverges.
\end{thm}

\begin{cor} \label{cor: consistency of eigencomponents of the second FPCA}
	Under the assumptions \ref{assmp: square int X}, \ref{assmp: reg con g(T)}, \ref{assmp: 4th moment Y} - \ref{assmp: jointly gaussian}, for each $k$ and $l$ the eigenvalues $\widehat \eta_{kl}$ and eigenfunctions $\widehat{\psi}_{kl}(\cdot) $ of $\widehat{G}_k(\cdot,\cdot)$ satisfy
	\begin{equation}
	|\widehat{\eta}_{kl} - \eta_{kl} | \xrightarrow{\text{  p  }} 0, \text{ and } \| \widehat{\psi}_{kl}(\cdot) - \psi_{kl}(\cdot) \|_{s} \xrightarrow{\text{  p  }} 0
	\end{equation}
	as $n$ diverges. Uniform convergence of $\widehat{\psi}_{kl}(\cdot)$ also holds:  $\text{sup}_{T}| \widehat{\psi}_{kl}(T) - \psi_{kl}(T) | \xrightarrow{\text{  p  }} 0$. Furthermore, as $n$ diverges, the estimator of the error variance, $\widehat{\sigma}^2_{e,k}$, and the functional principal component scores, $\widehat{\zeta}_{ikl}$'s, satisfy
	\begin{equation}
	|\widehat{\sigma}^2_{e,k}- \sigma^2_{e,k}|  \xrightarrow{\text{  p  }} 0
	\text{  and  }
	|\widehat{\zeta}_{ikl} - \widetilde{\zeta}_{ikl}| \xrightarrow{\text{  p  }} 0,
	\end{equation} 
where $\widetilde{\zeta}_{ikl} = {\text{E}[\zeta_{ikl}| {\boldsymbol{\xi}}_{W,ik} ]}$ and $ {\boldsymbol{\xi}}_{W,ik}$ is the $m_i$-dimensional column vector of ${\xi}_{W,ijk}$'s.
\end{cor}

The consistency results for all model components imply consistency of the predicted trajectories given in equation (\ref{eq: predicted trajectory}). 

\begin{conject} \label{th: consistency of predicted trajectories}
	Under the assumptions \ref{assmp: square int X}, \ref{assmp: reg con g(T)}, \ref{assmp: 4th moment Y} - \ref{assmp: jointly gaussian}, for each $(s,T) \in \mathcal{S} \times \mathcal{T}$
	\begin{equation}
	\widehat{Y}_{i}(s,T) \xrightarrow{\text{  p  }} \sum_{k=1}^{\infty} \sum_{l=1}^{\infty} \widetilde{\zeta}_{ikl} \psi_{kl}(T)\phi_k(s),
	\end{equation}
	as $n$, $K$ and $L_k$'s diverge.
\end{conject}

\subsection{Response curves measured with smooth error}  \label{subsec: theoretical results with error process}

Assume next that the response curves are observed with smooth error $ \epsilon_{ij}(s)\equiv \epsilon_{1,ij}(s)$, and thus $Y_{ij}(s) = X_{i}(s, T_{ij}) +\epsilon_{1,ij}(s)$ for $s\in \mathcal{S}$ and smooth error process $\epsilon_{1,ij}(\cdot)\in L^2(\mathcal{S})$.

The main difference from Section \ref{subsec: theoretical results with no errors} is that the sample covariance of $Y_{ij}(s)$ is an estimator of $\Xi(s,s')=\Sigma(s,s') + \Gamma(s,s')$, not of $\Sigma(s,s')$; as a result we denote the sample covariance of $Y_{ij}(s)$ by $\widehat{\Xi}(s,s')=\sum_{i=1}^{n} \sum_{j=1}^{m_i} Y_{ij}(s)Y_{ij}(s')/$ $(\sum_{i=1}^{n} m_i)$. One can show, using similar arguments as earlier, that $\widehat{\Xi}(s,s')$ is an unbiased estimator of the marginal covariance function, $\Xi(s,s')$. Moreover similar arguments can be used to show the pointwise consistency as well as the Hilbert-Schmidt norm consistency of $\widehat{\Xi}(s,s')$. 

\begin{enumerate} [resume, label= (A\arabic*.) ]
	\item Assume $\epsilon_{ij}(s)$ is realization of $\epsilon=\{\epsilon(s): s\in \mathcal{S} \}$, which is square integrable process in $L^2(\mathcal{S})$; recall $\epsilon_{ij}(s) = \epsilon_{1,ij}(s) + \epsilon_{2, ij}(s)$.\label{assmp: square int e1}
	\item $\text{E}\big[ \|\epsilon(\cdot) \|^4 \big] < \infty $. \label{assmp: 4th moment e1}
	\item $\text{E}\big[ \text{sup}_{s} |\epsilon(s)|^a \big] \leq M^a$ for a constant, $M > 0$, and an arbitrary integer, $a \geq 1$; this is equivalent to assume that $\epsilon(s)$ is absolutely bounded almost surely. \label{assmp: e1 asymptotically bounded}
\end{enumerate}

\begin{cor} \label{cor: Big Xi consistency}
	Under the assumptions \ref{assmp: square int X} - \ref{assmp: E[A(T)A(T')|T,T']}, and \ref{assmp: square int e1}, for each $(s,s')$, $
	|\widehat{\Xi}(s,s') - \Xi(s,s')| \xrightarrow{\text{  p  }} 0 $
	as $n$ diverges. And under the assumptions \ref{assmp: square int X}, \ref{assmp: reg con g(T)}, \ref{assmp: 4th moment Y}, \ref{assmp: square int e1}-\ref{assmp: 4th moment e1}, 
	$\| \widehat{\Xi}(\cdot, \cdot) - \Xi(\cdot, \cdot) \|_{s} \xrightarrow{\text{  p  }} 0$  and $\text{sup}_{j} | \widetilde{\xi}_{W,ijk} - \xi_{W,ijk}| \xrightarrow{\text{  p  }} 0 \text{ as } n \longrightarrow \infty$.
\end{cor}
The proofs of these results are detailed in the Supplementary Material.

As the smooth error process $\epsilon_{1,ij}(s)$ is correlated only along the functional argument, $s$, and  $\epsilon_{1,ij}(s)$ are iid over $i, j$ it follows that the theoretical properties of the predictions - of the time-varying coefficients and the response curve - hold without any modification.

\subsection{Extensions} \label{subsec: theoretical results discussion}

The theoretical results presented in this section are based on the assumption that data $Y_{ij}(s)$'s are observed (i) fully, (ii) without white noise, $\epsilon_{2,ij}(s)\equiv 0$ for all $s$, and (iii) have mean zero. These assumptions are made for convenience, and they can be relaxed as we now explain. 

(i) The assumption that $Y_{ij}(s)$'s are observed fully, as a continuous function, is quite common in theoretical study involving functional data; see for example \cite{cardot2003testing,cardot2004testing, chen2013repeated} among many others. One possibility to bypass this assumption is to use the corresponding smooth trajectories instead. (ii) Suppose that the profiles $Y_{ij}(\cdot)$'s are observed on dense grids of points and the measurements are additionally corrupted with white noise. \citet{zhang2007statistical} showed that by smoothing each profile using local linear smoother, the true de-noised curves are recovered with asymptotically negligible error, in the case of independent curves. Another possibility to handle white noise is to use ideas similar to \citet{yao2005functional}. In Section \ref{sec: simulation} we illustrate numerically the effect of white noise on the performance accuracy. (iii) Finally, the theoretical properties of the model components estimators remain valid, when the mean function is non-zero and a consistent mean estimator is available; \cite{chen2013repeated} had considered this problem and showed that under suitable assumptions such consistent mean estimator can be obtained using bivariate smoothing.

\section{Simulation study} \label{sec: simulation}

We conduct a simulation study to investigate the finite sample performance of the proposed modeling approach. The performance is evaluated in terms of estimation accuracy for the main model components, in-sample and out-of-sample prediction accuracy, and computational time. Whenever applicable we compare our results to available alternatives 
such as \citet{chen2013repeated}. 

We generate $N_{sim} = 1000$ samples from model (\ref{eq: modeling framework}) with $K=2$, $Y_{ij}(s) = \mu(s, T_{ij}) +  \xi_{i1}(T_{ij})   \phi_1(s) +  \xi_{i2}(T_{ij})   \phi_2(s)+ \epsilon_{ij}(s)$, where $\mu(s, T) = 1+2s+3T+4sT$, and $\phi_1(s)=1$ and $\phi_2(s) = \sqrt{2}\textrm{sin}(2\pi s)$. The grid of points for $s$ is the set of $101$ equispaced points in $[0,1]$. For each $i$, there are $m_i$ profiles associated with visit times, $\{T_{ij}: j=1, \ldots, m_i\}$; $T_{ij}$'s are randomly sampled from $41$ equally spaced points in $[0,1]$. The random coefficients $\xi_{ik}(T)$ are generated from various covariance structures as detailed below. Errors are generated from $\epsilon_{ij}(s) = e_{ij1} \phi_1(s)+ e_{ij2}\phi_2(s) + \epsilon_{2,ij}(s)$, where  $e_{ij1}$, $e_{ij2}$ and $\epsilon_{2,ij}(s)$ are mutually independent with zero-mean and variances equal to $\sigma^2_{e, 1}, \sigma^2_{e, 2}$ and $\sigma^2$, respectively. The white noise variance, $\sigma^2$, is set based on the signal to noise ratio (SNR), 
\begin{equation}
\textrm{SNR} = \dfrac{\iint \textrm{var}\{Y_{i}(s,T)\} ds dT}{(\sigma^2_{e,1} + \sigma^2_{e,2} + \sigma^2)} - 1 .
\end{equation}
 We consider the following experimental factors:
 
\begin{enumerate}[label= Case \arabic{*}.]
\item covariance structure of the time-varying components: \\
(a) non-parametric covariance (NP): $\xi_{ik}(T) = \zeta_{ik1} \psi_{k1}(T)+  \zeta_{ik2} \psi_{k2}(T)$, where \\
(i) $\psi_{11}(T) = \sqrt{2}\textrm{cos}(2\pi T)$, $\psi_{12}(T) = \sqrt{2}\textrm{sin}(2\pi T)$, $\zeta_{i11} \overset{\textrm{iid}} \sim \textrm{N}(0,3)$, $\zeta_{i12} \overset{\textrm{iid}} \sim \textrm{N}(0,1.5)$; \\
(ii) $\psi_{21}(T) = \sqrt{2}\textrm{cos}(4\pi T)$, $\psi_{22}(T) = \sqrt{2}\textrm{sin}(4\pi T)$, $\zeta_{i21} \overset{\textrm{iid}} \sim \textrm{N}(0,2)$, $\zeta_{i22} \overset{\textrm{iid}} \sim \textrm{N}(0,1)$.\\
(b) random effects model (REM): $\xi_{ik}(T) = b_{ik0} + b_{ik1}T$ with 
\begin{eqnarray*}
\begin{pmatrix}b_{i10}\\
b_{i11}
\end{pmatrix}  \overset{\textrm{iid}} \sim N\left[\left(\begin{array}{c}
0\\
0
\end{array}\right),\left(\begin{array}{cc}
2.5 & 2 \\
2 & 3
\end{array}\right)\right],
\begin{pmatrix}b_{i20}\\
b_{i21}
\end{pmatrix} \overset{\textrm{iid}} \sim   N\left[\left(\begin{array}{c}
0\\
0
\end{array}\right),\left(\begin{array}{cc}
2 & 1\\
1& 1.5
\end{array}\right)\right],
\end{eqnarray*}
(c) exponential autocorrelation model (Exp): $\xi_{ik}(T)$ is a Gaussian process with mean zero, variance $\lambda_k$ and auto-correlation function $\textrm{corr}\{\xi_{ik}(T), \xi_{ik}(T')\} = \rho_k^{|T-T'|}$ denoted by $GP(\lambda_k, \rho_k)$. We set $\xi_{i1}(T) \overset{\textrm{iid}} \sim GP(4.5, 0.9)$ and $\xi_{i2}(T)  \overset{\textrm{iid}}\sim GP(3, 0.5)$.

Note that regardless of the generating models for $\xi_{ik}(T)$, we have that $\int \textrm{var}\{\xi_{ik}(T)\}dT$ is equal to $4.5$ and $3$ for $k=1,2$ respectively.	
	
\item number of repeated measurements per subject: \\
(a) $m_i \overset{iid} \sim \textrm{Uniform}(\{8, 9,\ldots, 12\})$ (about $75\%$ missing)\\
(b) $m_i \overset{iid} \sim \textrm{Uniform}(\{15, 16, \ldots, 20\})$ (about $55\%$ missing)

\item variance of $e_{ijk}$:\\
(a) $(\sigma^2_{e, 1}, \sigma^2_{e, 2}) = (0,0)$ (white noise only, i.e. $\epsilon_{ij}(s) \overset{\textrm{iid}} \sim N(0, \sigma^2)$)\\
(b) $(\sigma^2_{e, 1}, \sigma^2_{e, 2}) = (0.7,0.3)$. 
\\\\
The simulation results for the Case 3.(a), i.e. no smooth error, are included in the Supplementary Material.

\item signal to noise ratio: (a) $ \textrm{SNR}=1 \; (\sigma^2=6.5)$, and (b) $ \textrm{SNR}=5 \; (\sigma^2=0.5)$

\item number of subjects: (a) $n=100$, (b) $n=300$, and (c) $n=500$
\end{enumerate}

For each generated sample of size $n$ we form a training set and a test set. To determine the test set we randomly select $10$ subjects from the sample. The test set is formed by collecting these subjects' last functional observation; hence the test set contains $10$ curves. The remaining functional observations for the $10$ subjects and the data corresponding to the remaining subjects in the sample form the training set. 
Our model is fitted using the training set and the methods outlined in Section \ref{sec: estimation}. To be more specific, the bivariate mean function, $\mu(s, T)$, is modeled using $50$ cubic spline basis functions obtained from the tensor product of $d_s=10$ basis functions in direction $s$ and $d_T = 5$ in $T$. The smoothing parameters are selected via REML. The finite truncations $K$ and $L_k$'s are all estimated using the pre-specified level PVE = $0.95$.

Estimation accuracy for the model components is evaluated using integrated mean squared errors (IMSE): specifically, for the bivariate mean function $\textrm{IMSE}(\widehat{\mu}) = \sum_{i_{sim}=1}^{N_{sim}} $ $\iint  \{\widehat{\mu}^{i_{sim}}(s,T)$ $- \mu(s,T)\}^2 dsdT/ N_{sim}$, and for the univariate eigenfunctions $\textrm{IMSE}(\widehat{\phi_k}) = \sum_{i_{sim}=1}^{N_{sim}}$ $ \int  \{\widehat{\phi}_{k}^{i_{sim}}(s) - \phi_{k}(s)\}^2  ds/  N_{sim}$, $k=1,2$. %
The prediction performance is assessed through the accuracy in predicting the time-varying model components, $\xi_{ik}(T)$, and in predicting the response curve, $Y_i(s,T)$. For the former assessment we use the in-sample integrated prediction errors (IPE) defined as $\textrm{IPE}(\xi_{k}) = \sum_{i_{sim}=1}^{N_{sim}}  \sum_{i=1}^{n} \int \{\widehat{\xi}_{ik}^{i_{sim}}(T) - \xi_{ik}^{i_{sim}}(T)\}^2dT /$ $(nN_{sim}) $, $k=1,2$. For the later assessment we use the in-sample IPE (IN-IPE) defined as $\textrm{IN-IPE}(Y)= \sum_{i_{sim}=1}^{N_{sim}} \sum_{i}^n\sum_{j=1}^{m_i} \int  \{\widehat{Y}_{ij}^{ i_{sim}}(s) - Y_{ij}^{*, i_{sim}}(s)\}^2ds /(N_{sim}$ $\sum_{i=1}^n m_i )$, where $Y^{*}_{ij}(s)$ is the true signal, i.e. without measurement error $\epsilon_{ij}(s)$. Also we use the out-of-sample IPE (OUT-IPE) defined as $\textrm{OUT-IPE}(Y)= \sum_{i_{sim}=1}^{N_{sim}} \sum_{i\in \text{test set} } \int  \{\widehat{Y}_{im_i}^{ i_{sim}}(s) - Y_{im_i}^{*, i_{sim}}(s)\}^2ds /(10  N_{sim})$ and $Y^{*}_{ij}(s)$ is the true signal in the test set. The results are based on $N_{sim}=1000$ simulations.

In terms of estimation performance and prediction of $\xi_{ik}(T)$ there is no alternative approach. On the other hand, in terms of model prediction error and prediction of a subject's future curve there are two possible alternatives. One is the CM model of \citet{chen2013repeated}. However due to the high computational expense required by their method, we have to restrict our comparison to few scenarios only:  $m_i \sim \{8, \ldots, 12\}$ number of repeated curves per subject, Case 3(b), and $\text{SNR} = 1$. The approach of \citet{chen2013repeated} requires specification of several kernel bandwidths; due to the increased computation burden we use the pre-specified bandwidth $h=0.1$ in smoothing both the mean and covariance functions. Even with these adjustments there is an order of magnitude difference in the computational cost (when $n=100$ the method of \citet{chen2013repeated} takes approximately $984$ seconds, while our approach takes about $7$ seconds). As well, we also used the pre-specified level $\text{PVE}=0.95$ to be consistent with our approach. A second alternative approach for prediction of a subject's future visit trajectory is a rather na\"ive approach: let the future prediction equal the average of all previously observed profiles for that subject. For example, the na\"ive predictor of a profile of some subject in the test set is equal to the average of all profiles available in the training set for the corresponding subject.

Table \ref{tab: functional error processes} shows the results for different covariance models for $\xi_{ik}(T)$, different number of repeated curve measurements per subject, different SNRs, complex error process, and varying sample sizes; basically the results for Case 1 (a)-(c), Case 2(a)-(b), Case 3 (b), Case 4 (a)-(b), and Case 5 (a)-(c). The analogous results corresponding to trivial covariance structure of the error process (white noise) are included in Table S1 of the Supplementary Material. The performance of the proposed estimation (see columns for $\mu$, $\phi_1$, and $\phi_2$ of this table) is slightly affected by the covariance structure of model components describing the dynamic behavior, $\xi_{ik}(T)$'s, and the number of repeated curve measurements per subject, but in general is quite robust to the factors we investigated. As expected the estimation accuracy improves with larger sample size; see the $3\times 3$ top left block of IMSE results corresponding to $n=100$, $n=200$, and $n=500$. The results corresponding to white noise measurement error are consistent with these observations. 

To describe the prediction performance we consider both the prediction of $\xi_{ik}(T)$'s and the prediction of the curve responses; consider columns labeled $\xi_1$, $\xi_2$, IN-IPE and OUT-IPE of Table \ref{tab: functional error processes}. As expected the underlying covariance structure of $\xi_{ik}(T)$'s does affect the prediction accuracy. In our investigation it seems that the exponential covariance structure (Exp) is most challenging; this most likely is due to the (very) large correlation coefficients used $\rho_1=0.9$ and $\rho_2=0.5$, which result in high temporal dependence even for the observations that are furthest apart $T=0$ and $T=1$, as $T\in [0,1]$. Increasing the level of signal relative to the magnitude of noise (SNR) does improve the results somewhat. More importantly, when the error process has trivial covariance structure, the prediction accuracy is greatly improved. The results also show an interesting finding: increasing the number of repeated curve measurements $m_i$ has a greater effect on the accuracy than increasing the sample size $n$. This observation should not be surprising, as with larger number of repeated measurements the estimation of the covariance of the longitudinal process $\xi_{ik}(T)$'s improves and as a result superior prediction.    
 
Here is a summary of the comparison between the proposed method and available alternatives. As already pointed out, the comparison with CM is limited by the computational expense involved. In interest of space the results are presented in Table S4 of the Supplementary Material: they show that the prediction using CM is more sensitive to the covariance structure of the underlying time-varying coefficients and that the accuracy can be improved by up to $50\%$ by the proposed approach. As expected, the na\"ive approach (results presented in the columns labeled $\text{IN-IPE}_\text{naive}$ and $\text{OUT-IPE}_\text{naive}$ of Table \ref{tab: functional error processes}) is very sensitive to the covariance structure of the latent longitudinal components, $\xi_{ik}(T)$. When the correlation structure is simple (random effects model, REM, and exponential dependence, Exp) it yields much better results compared to when the correlation is more complex (nonparametric, NP). Not surprisingly, in all the cases studied the prediction accuracy is inferior to the proposed method. Table \ref{tab: time elapsedP} provides insights into the how the computational time of our method scales with the number of subjects; for completeness the computational time is studied for all the cases in Tables S2 and S3 of the Supplementary Material. 

The overall conclusion is that the proposed approach provides an improved prediction performance over the existing methods in a computationally efficient manner.

\begin{table}[htbp] 
\caption {Estimation and prediction accuracy results based on $N_{sim}=1000$ simulations} \label{tab: functional error processes}
\centering
\scalebox{0.63}{
\begin{tabular}{rr|ccccccccc}
  \midrule
  \multicolumn{11}{c}{$m_i \sim \{8, \ldots, 12\} \textrm{ and SNR}= 1$} \\ \hline
& & $\mu$ & $\phi_1$ & $\phi_2$ & $\xi_1$ & $\xi_2$ & IN-IPE & IN-IPE$_{\textrm{naive}}$ & OUT-IPE & OUT-IPE$_{\textrm{naive}}$\\ 
  \hline

NP (a) & $n=100$ &  0.092 & 0.003 & 0.011 & 0.338 & 0.224 & 0.406 & 7.790 & 0.988 & 11.478 \\ 
& $n=300$ & 0.031 & 0.001 & 0.009 & 0.226 & 0.138 & 0.313 & 7.773 & 0.559 & 11.349 \\ 
& $n=500$ & 0.019 & 0.001 & 0.009 & 0.199 & 0.117 & 0.288 & 7.779 & 0.455 & 11.262 \\ 
REM (b) & $n=100$ &  0.114 & 0.027 & 0.033 & 0.376 & 0.314 & 0.328 & 1.199 & 1.011 & 2.160 \\ 
& $n=300$ & 0.040 & 0.008 & 0.013 & 0.216 & 0.162 & 0.265 & 1.197 & 0.675 & 2.160 \\ 
& $n=500$ & 0.024 & 0.005 & 0.010 & 0.181 & 0.133 & 0.247 & 1.197 & 0.571 & 2.150 \\ 
Exp (c) & $n=100$ &  0.095 & 0.022 & 0.030 & 0.399 & 0.540 & 0.554 & 1.528 & 1.426 & 2.520 \\ 
& $n=300$ & 0.031 & 0.007 & 0.015 & 0.289 & 0.412 & 0.508 & 1.531 & 1.143 & 2.498 \\ 
& $n=500$ & 0.019 & 0.004 & 0.013 & 0.266 & 0.383 & 0.494 & 1.530 & 1.074 & 2.492 \\ 
\midrule
  
\multicolumn{11}{c}{$m_i \sim \{15, \ldots, 20\} \textrm{ and SNR}= 1$}   \\ \hline
& & $\mu$ & $\phi_1$ & $\phi_2$ & $\xi_1$ & $\xi_2$ & IN-IPE & IN-IPE$_{\textrm{naive}}$ & OUT-IPE & OUT-IPE$_{\textrm{naive}}$ \\ 
  \hline
NP (a) & $n=100$ &  0.076 & 0.002 & 0.010 & 0.180 & 0.101 & 0.238 & 7.807 & 0.477 & 10.666 \\ 
& $n=300$ & 0.026 & $<0.001$ & 0.009 & 0.120 & 0.065 & 0.183 & 7.796 & 0.282 & 10.728 \\ 
& $n=500$ & 0.016 & $<0.001$ & 0.009 & 0.108 & 0.058 & 0.173 & 7.797 & 0.242 & 10.772 \\ 
REM (b) & $n=100$ &  0.097 & 0.025 & 0.031 & 0.272 & 0.252 & 0.232 & 0.897 & 0.612 & 1.833 \\ 
& $n=300$ & 0.034 & 0.008 & 0.013 & 0.156 & 0.132 & 0.201 & 0.896 & 0.462 & 1.841 \\ 
& $n=500$ &0.020 & 0.005 & 0.010 & 0.135 & 0.110 & 0.194 & 0.897 & 0.440 & 1.836 \\ 
Exp (c) & $n=100$ &  0.080 & 0.022 & 0.030 & 0.308 & 0.417 & 0.467 & 1.240 & 1.048 & 2.147 \\ 
& $n=300$ & 0.026 & 0.006 & 0.015 & 0.233 & 0.309 & 0.444 & 1.245 & 0.938 & 2.155 \\ 
& $n=500$ & 0.016 & 0.004 & 0.012 & 0.221 & 0.285 & 0.438 & 1.246 & 0.886 & 2.129 \\ 
\midrule

\multicolumn{11}{c}{$m_i \sim \{8, \ldots, 12\} \textrm{ and SNR}= 5$}   \\ \hline
& & $\mu$ & $\phi_1$ & $\phi_2$ & $\xi_1$ & $\xi_2$ & IN-IPE & IN-IPE$_{\textrm{naive}}$ & OUT-IPE & OUT-IPE$_{\textrm{naive}}$\\ 
  \hline
NP (a) & $n=100$ & 0.092 & 0.005 & 0.005 & 0.328 & 0.213 & 0.363 & 7.184 & 0.958 & 10.795 \\ 
& $n=300$ & 0.031 & 0.001 & 0.002 & 0.213 & 0.124 & 0.268 & 7.170 & 0.506 & 10.662 \\ 
& $n=500$ & 0.019 & 0.001 & 0.001 & 0.187 & 0.103 & 0.242 & 7.178 & 0.402 & 10.585 \\ 
REM (b) & $n=100$ & 0.114 & 0.037 & 0.037 & 0.404 & 0.355 & 0.293 & 0.594 & 0.958 & 1.478 \\ 
& $n=300$ &0.040 & 0.010 & 0.011 & 0.218 & 0.167 & 0.235 & 0.595 & 0.627 & 1.476 \\ 
& $n=500$ & 0.024 & 0.006 & 0.007 & 0.180 & 0.135 & 0.219 & 0.596 & 0.529 & 1.467 \\ 
Exp (c) & $n=100$ &0.095 & 0.033 & 0.033 & 0.420 & 0.573 & 0.513 & 0.922 & 1.419 & 1.838 \\ 
& $n=300$ &0.031 & 0.010 & 0.010 & 0.290 & 0.412 & 0.466 & 0.929 & 1.109 & 1.814 \\ 
& $n=500$ &0.019 & 0.006 & 0.006 & 0.264 & 0.378 & 0.453 & 0.929 & 1.033 & 1.807 \\ 

\midrule
\multicolumn{11}{c}{$m_i \sim \{15, \ldots, 20\} \textrm{ and SNR}= 5$}   \\ \hline
& & $\mu$ & $\phi_1$ & $\phi_2$ & $\xi_1$ & $\xi_2$  & IN-IPE & IN-IPE$_{\textrm{naive}}$ & OUT-IPE & OUT-IPE$_{\textrm{naive}}$\\  
  \hline  
NP (a) & $n=100$ &  0.076 & 0.003 & 0.003 & 0.174 & 0.095 & 0.205 & 7.462 & 0.441 & 10.300 \\ 
 & $n=300$ & 0.026 & 0.001 & 0.001 & 0.113 & 0.057 & 0.147 & 7.453 & 0.239 & 10.359 \\ 
 & $n=500$ & 0.016 & $<0.001$ & 0.001 & 0.101 & 0.050 & 0.136 & 7.454 & 0.200 & 10.406 \\ 
REM (b) & $n=100$ &  0.097 & 0.035 & 0.035 & 0.300 & 0.293 & 0.205 & 0.552 & 0.568 & 1.464 \\ 
 & $n=300$ & 0.034 & 0.010 & 0.010 & 0.160 & 0.140 & 0.178 & 0.552 & 0.426 & 1.473 \\ 
 & $n=500$ & 0.020 & 0.006 & 0.007 & 0.136 & 0.114 & 0.172 & 0.554 & 0.405 & 1.470 \\ 
Exp (c) & $n=100$ &  0.080 & 0.033 & 0.033 & 0.330 & 0.451 & 0.434 & 0.895 & 1.012 & 1.779 \\ 
 & $n=300$ & 0.027 & 0.009 & 0.010 & 0.236 & 0.313 & 0.410 & 0.902 & 0.901 & 1.785 \\ 
 & $n=500$ & 0.016 & 0.005 & 0.006 & 0.221 & 0.284 & 0.403 & 0.902 & 0.851 & 1.763 \\   
  \midrule
\end{tabular}
}
\end{table}

\begin{table}[htbp]
\caption {Computational time (seconds)} \label{tab: time elapsedP}
\centering
\scalebox{.63}{
\begin{tabular}{r|ccc}
    \midrule
\multicolumn{4}{c}{$m_i \sim \{8, \ldots, 12\} \textrm{ and SNR}= 1$}   \\ \hline
 & $n=100$ & $n=300$ & $n=500$ \\ \hline
NP (a)  & 7.369 & 15.892 & 21.418 \\ 
REM (b)  & 9.282 & 11.347 & 22.559 \\ 
Exp (c)  & 7.514 & 16.229 & 17.109 \\ 
   \hline
\end{tabular}
}
\end{table}

\section{DTI application} \label{sec: application}

Diffusion tensor imaging (DTI) is a magnetic resonance imaging technique, which is crucial in the diagnosis and progression assessment of various neuro-pathological diseases that affect brain white matter. Specifically, DTI provides different measures of water diffusivity along brain white matter tracts;
its use is instrumental especially for brain tracts where the diffusion is sensitive to any alteration of the tissues in the tract, like the ones caused by multiple sclerosis (MS), or white matter diseases more generally (see \cite{alexander2007diffusion}, \cite{basser1994mr}, \cite{basser2000vivo}, \cite{basser2011microstructural}).

In this paper we focus on studying a commonly used DTI measure - fractional anisotropy (FA) - along the corpus callosum tract in MS patients. FA of the water diffusion ranges from zero to one, with zero being the perfect isotropic diffusion in all directions. The DTI study involves $162$ MS patients observed at between one and eight hospital visits; the total number of visits for all patients is $421$ with mean and median equal to $2.599$ and $2$ visits per patient, respectively. At each visit, FA is measured at $93$ equispaced discrete locations along the corpus callosum tract, and thus can be viewed as a typical densely-sampled functional data. The data contains a total of $39$ missing values; however, no modification is needed as the proposed method is not sensitive to mild missingness. 

Our main objective is twofold: (i) to understand the dynamic behavior of the FA profile in MS patients over time and (ii) to make accurate predictions of the FA profile for the observed subject's future visit. Various aspects of the DTI study have been also considered in \citet{greven2010}, \citet{goldsmith2011penalized}, \citet{staicu2012modeling} and \citet{pomann2013two}. In particular \citet{greven2010} used an earlier version of the study consisting of data from fewer and possibly different
patients and a different registration technique. They studied the dynamic behavior of FA over time; however, their method cannot provide prediction for an already observed patient of the FA profile at their next hospital visit. Our method quantifies the longitudinal changes in FA over time using a parsimonious model framework and provides accurate prediction of the full response profile at patient's future visit. The results have the potential to shed lights on the understanding of the MS progression over time as well as its response to treatment. 

To start with, for each subject we define the hospital visit time $T_{ij}$ by the difference between the reported visit time and the subject's baseline visit; thus $T_{i1}=0$ for all subjects $i$. Also the resulting values are scaled by the maximum value in the study so that $T_{ij}\in  [0,1]$ for all $i$ and $j$. 
The sampling distribution of the visit times is right-skewed with rather strong skewness; for example there are only few observations $T_{ij}$'s close to $1$. The strong skewness of the sampling distribution of $T_{ij}$'s has serious implications on the estimation of the bivariate mean $\mu(s,T)$; a completely nonparametric bivariate smoothing would results in unstable and highly variable estimation. This is probably why \citet{greven2010} first centered the times for each patient $i$, $\{T_{ij}: j=1, \ldots, m_i\}$, and then standardized the overall set $\{T_{ij}:i,j\}$ to have unit variance. However, such subject-specific transformation
of $T_{ij}$'s loses interpretability. In particular, while such transformations are appropriate for understanding the complex dynamics of these data they may not be suited for prediction at unobserved times. Specifically, it is not clear what type of transformation to apply to a future visit time of a subject, in order to predict the response profile at the respective future time - and this is crucial for our analysis.

\begin{figure}[h!]
\caption{{Left panel:} $95\%$ pointwise and joint confidence bands of the slope function $\beta_T(s)$ of $\mu(s,T)$ using bootstrap; {Right}: final mean estimate, $\hat{\mu}(s,T)=\hat{\mu}_0(s)$}
\centering
\includegraphics[width=1\textwidth]{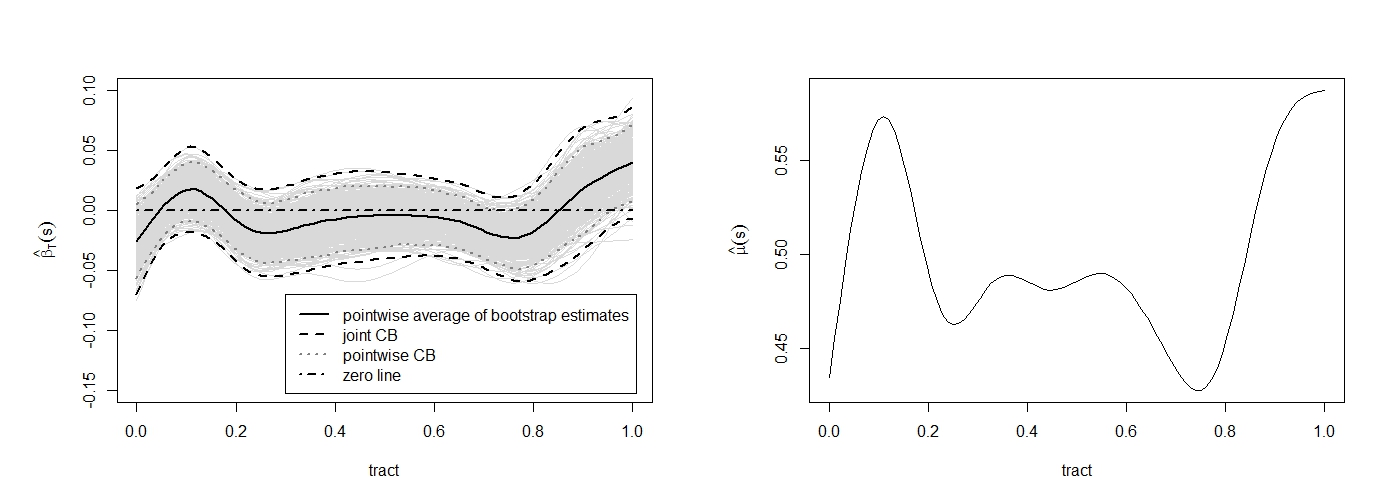}
\label{fig: mean coef bootstrap}
\end{figure}

One way to bypass this challenge is to assume a simpler parametric structure along the longitudinal direction, $T$, for the mean function; based on the exploratory analysis we assume linearity in $T$. Specifically we assume the varying coefficient model $\mu(s,T_{ij}) = \mu_0(s) + \beta_{T}(s)T_{ij}$, where $\mu_0(\cdot)$ and $\beta_{T}(\cdot)$ are unknown, smooth functions of $s$. We model $\mu_0(\cdot)$ and $\beta_T(\cdot)$ using a penalized univariate cubic spline regression with $10$ basis functions; the smoothing parameters are estimated using REML. Plots of the estimated mean, $\widehat{\mu}(s,T)= \widehat{\mu}_0(s) + \widehat{\beta}_{T}(s) T$, and the estimated slope function, $\widehat{\beta}_{T}(s)$, are given in Figure S1 of the Supplementary Material. The estimates seem to indicate that the population-level mean of FA, $\widehat{\mu}(s,T)$, does not change much over longitudinal time, $T$; in other words the estimated slope function, $\widehat{\beta}_{T}(s)$, is very close to zero. We use bootstrapping of the subjects to quantify the variability of the slope function estimator. In particular, pointwise confidence band is constructed as $2.5\%$ and $97.5\%$ quantiles of the bootstrap estimates at each location of the tract, $s$, and the joint confidence band is constructed as given in \citet{crainiceanu2011statistical} using $B=1000$ bootstrap samples. As shown in Figure \ref{fig: mean coef bootstrap} the joint confidence band contains zero for all $s$.

The observation that the mean of FA profile varies very little over time is in agreement with prior literature; for example see \citet{greven2010} who used a different data set with subjects observed also over a relatively short time frame. To gain more insight into this direction we consider a formal test to see whether $\mu(s,T)$ varies over time; using our mean model assumption this is equivalent to testing the null hypothesis that the slope function is null, $\text{H}_0: \beta_T(s) = 0$ against the alternative $\text{H}_{A}: \beta_T(s) \neq 0$ for some $s\in [0,1]$. We inspire from the ideas discussed in \citet{park2015fixed} to carry out the testing procedure.

Let the test statistic be $Q = \int \widehat{\beta}_T(s)^2 ds$ - the norm of the slope estimator. To approximate its null distribution, we use bootstrap of the residuals as described next. First, estimate the mean function by assuming the model $\mu(s,T)= {\mu}_0(s) + {\beta}_{T}(s) T$ and under working independence; let $Q_{obs} = \int \widehat{\beta}_T(s)^2 ds$ be the observed test statistic using numerical integration. Second, construct the bootstrap sample as $Y_{ij}^{\text{boot}}(s) = Y_{ij}(s) - \widehat \beta_T(s) T_{ij}$. Third, obtain $B$ bootstrap datasets, by bootstrapping with replacement from the above set; take $B=1000$. For each data set we fit the above mean model and calculate the bootstrap test statistic, denoted by $Q_{b}$, where $b$ indexes for bootstrap dataset. Fourth, approximate p-value by $\sum_{b=1}^{1000} I(Q_{b} > Q_{obs})/1000$, where $I$ is an indicator function. The null distribution of $Q$ is shown in Figure S2 of the Supplementary Material. The procedure yields p-value $=0.206$, which does not provide significant evidence to reject the null hypothesis that the mean function is constant over time. In light of this, we assume a $T$-invariant mean model $\mu(s,T) = \mu_0(s)$ for the rest of the data analysis; the estimated $\widehat \mu_0(s)$ is shown in Figure \ref{fig: mean coef bootstrap}.

\begin{figure}[h!]
  \caption{ {Top:} First three eigenfunctions of the estimated marginal covariance; {Bottom:} estimated mean function $\widehat{\mu}_0(s)$ (gray line) $\pm \; 2\sqrt{\widehat{\lambda}_k}\widehat{\phi}_k(s)$ ($+$ and $-$ signs, respectively)}
  \centering
  \includegraphics[width=1\textwidth]{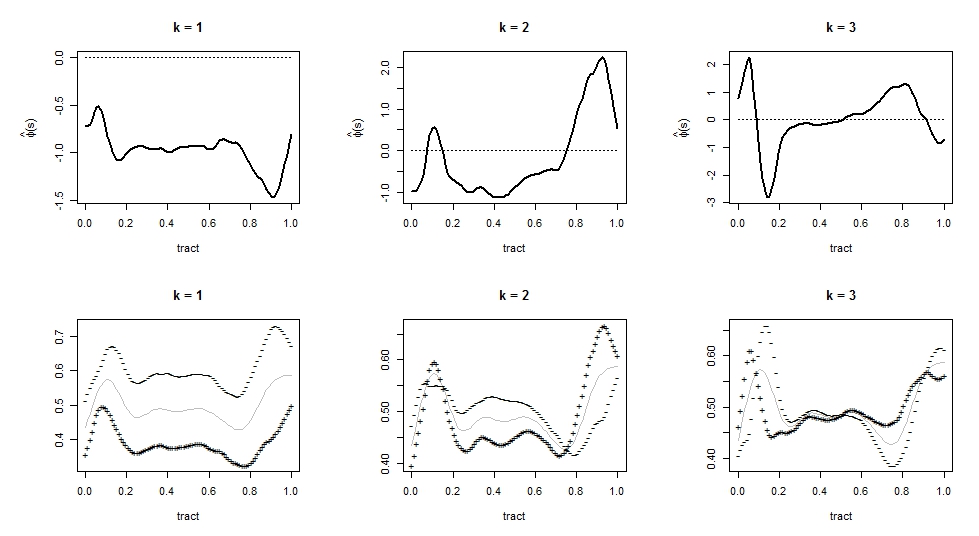}
  \label{fig: MFPCA eigenfunctions}
\end{figure}

We focus next on studying the variability in the data, and on using this variability to inform future prediction of a subject's full trajectory. After demeaning the data, we estimate the marginal covariance and obtain the eigenbasis functions; using the preset level $PVE=0.95 $ results in $K=10$ eigenfunctions. Figure \ref{fig: MFPCA eigenfunctions} shows the leading $3$ eigenfunctions that explain in turn $62.69\%$, $8.37\%$ and  $6.77\%$ of the total variance, respectively; the rest of the estimated eigenfunctions are given in Figure S4 of the Supplementary Material. As shown in Figure \ref{fig: MFPCA eigenfunctions}, a positively loaded first eigenfunction corresponds to a mean profile that is lower than the overall mean for all $s$, and the opposite for a negatively loaded one.

We initially use a nonparametric approach to estimate the longitudinal covariance of $\xi_{ik}(T)$'s for $k = 1,\dots, 10$. Preliminary results (not shown here) indicate a simpler covariance model, which is the model we will present. More specifically, assume a random effects model $\xi_{ik}(T_{ij}) = b_{0ik} + b_{1ik}T_{ij}$, where $\textrm{var}(b_{lik}) = \sigma^2_{lk}$ for $l=0,1$ and $\textrm{cov}(b_{0ik}, b_{1ik}) = \sigma_{01k}$. The fitted time-varying coefficient functions, $\widehat{\xi}_{ik}(T)$, for $k=1,2 \text{ and }3$ are shown in Figure \ref{fig: xi hat}, and the rest are shown in Figure S5 of the Supplementary Material. The estimated time-varying coefficient functions corresponding to $\widehat{\xi}_{i1}(T)$ suggest some longitudinal changes, but their signs remain the same across visit times except few cases with a relatively small magnitude. Overall, it implies that an individual mean profile tends to stay lower than the overall mean across all visit times if the first eigenfunction corresponding to that individual is positively loaded at baseline, and vise versa. In contrast to $\widehat{\xi}_{i1}(T)$, the estimated basis coefficient functions corresponding to the second eigenfunction, $\widehat{\xi}_{i2}(T)$, are mostly constant across visit times and imply little changes over time. 

One advantage of using a simpler, parametric model is that one can consider more formal testing about longitudinal effects. We consider this idea next, and study separately the null hypotheses, $\text{H}_{0k}: \sigma^2_{1k} =\sigma_{01k}= 0$ for $k = 1, \ldots, 10$ using the proxy data $\xi_{W,ijk}=\int \{Y_{ij}(s) -\widehat \mu_0(s) \} \widehat \phi_k(s)$ as the `observed' data. Bonferroni correction is used to appropriately account for multiple testings; specifically, we use the adjusted significance level, $\alpha_{adj} = 0.05/10 = 0.005$, for testing each hypothesis such that familywise error rate is $\alpha = 0.05$. Likelihood ratio test (LRT) and its null distribution as approximated by the mixture chi-squares are used for each hypothesis, $\text{H}_{0k}$. With the adjusted significance level, $\alpha_{adj}$, we reject the null hypotheses, $\text{H}_{0k}$, for $k=1,3,4, \text{ and }5$, with p-values less than $0.001$. We conclude that there is nontrivial longitudinal dynamics of FA. To the best of our knowledge, this is the first attempt to carry out a formal testing for longitudinal changes in functional observations over time. 

\begin{figure}[h!]
  \caption{Estimated time-varying coefficients $\widehat{\xi}_{ik}(T)$ for $k=1,2 \text{ and }3$ using REM}
  \centering
  \includegraphics[width=1\textwidth]{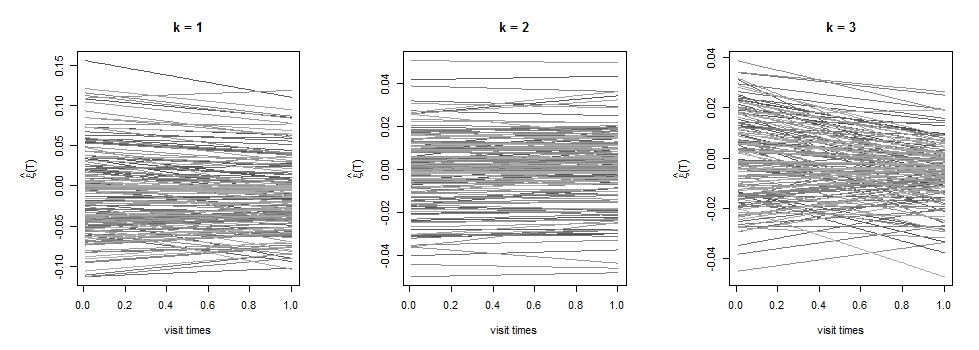}
  \label{fig: xi hat}
\end{figure}

Finally, we assess the goodness-of-fit and prediction accuracy of our final model. For the goodness-of-fit we use the in-sample integrated prediction error (IPE): IPE= $\sum_{i=1}^{162} \sum_{j=1}^{m_i}$ $\int \{ Y_{ij}(s) - \widehat{Y}_{ij}(s) \}^2 ds / \{\sum_{i=1}^{162} m_i\}  $, where $\widehat Y_{ij}(s)$ $= \widehat \mu_0(s) + \sum_{k=1}^{K} (\widehat b_{0ik} + \widehat b_{1ik}T_{ij}) \widehat \phi_k(s)$, and $Y_{ij}(s)$'s are the observed curve data. We compare our results with two other competitive approaches: \citet{greven2010} and \citet{chen2013repeated}. The square root of the in-sample IPEs are $2.31 \times 10^{-2}$ for our model, $2.66 \times 10^{-2}$ for \citet{greven2010}, and $3.76 \times 10^{-2}$ for \citet{chen2013repeated}. For all of three approaches we use the same mean function estimate, $\hat{\mu}(s,T)=\hat{\mu}_{0}(s)$, and the same pre-specified level $\text{PVE}=0.95$; for \citet{chen2013repeated} we use the pre-set bandwidth $h= 0.1$.

\begin{figure}[h!]
  \caption{Predicted values of FA for the last visits of three randomly selected subjects; actual observations (gray); predictions using our model (black solid) and using the naive approach (black dashed)}
  \centering
    \includegraphics[width=1\textwidth]{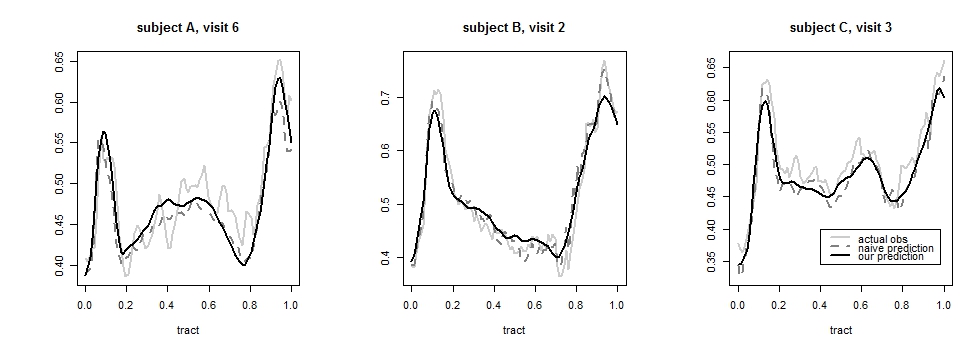}
    \label{fig: predicted}
\end{figure}

To assess the prediction accuracy we use leave-the last-curve-out integrated prediction error: $ \sum_{i=1}^{106}  \int  \{ Y_{im_i}(s) - \widehat{Y}^{[-i m_i]}_{im_i}(s) \}^2ds / 106  $, where $\widehat{Y}^{[-i m_i]}_{im_i}(s)$ is the predicted curve for the $i$th subject using the fitted model based on all the data less the $m_i$th curve of the $i$th subject. Specifically, for each of $106$ patients with more than one hospital visits, we remove the FA measurements taken at the last hospital visit from the dataset; the obtained dataset is used to fit the proposed model and predict the removed curve measurements. Figure \ref{fig: predicted} shows such predicted curves $\widehat{Y}^{[-im_i]}_{im_i}(s)$ obtained using our model and the naive model for three randomly selected subjects' FAs at their last visits. The alternative methods are \citet{chen2013repeated} and the naive approach described in Section \ref{sec: simulation}. The square root of the leave-the last-curve-out integrated prediction errors are $3.48 \times 10^{-2}$ for our model,  $8.71 \times 10^{-2}$ for \citet{chen2013repeated}, and $3.52 \times 10^{-2}$ for the na\"ive approach. Surprisingly, the naive estimation has relatively good prediction performance, and better than \citet{chen2013repeated}. Nevertheless these results confirm that, in this short term study of MS, there is a small variation of FAs within subjects over time.

\section{Discussion} \label{sec: discussion}

In this paper we propose a novel parsimonious modeling framework for repeatedly observed functional
data, due to a longitudinal design. Accounting for the dependence within the subject as well as for the longitudinal design is crucial for making full prediction at future visits, or assessing the dynamics of the underlying process
over time. However, current methods either ignore the dependence or are too complicated and
computationally intensive. 

Using Shiny \citep{shiny} we implemented interactive plots to help visualize longitudinal functional data as well as the various components and prediction obtained using the proposed method; these tools will shortly be accessible on {\tt GitHub}. {\tt R} code for illustrating the procedure is publicly available at {\url{http://www4.stat.ncsu.edu/~staicu/software/MLFD_Rcode.zip}  }


%
%
\setcounter{section}{1}
\renewcommand{\thesection}{A\arabic{section}}

\section*{Appendix}
\subsection*{Proof for Theorem \ref{th: consistency of Sigma}}

We prove this theorem by first proving the pointwise consistency and then proving the Hilbert-Schmidt norm consistency of the covariance estimator.\\

\noindent \emph{Pointwise consistency}:\\

Recall $\widehat{\Sigma}(s,s') = \sum_{i=1}^n \sum_{j=1}^{m_i} Y_{ij}(s) Y_{ij}(s') /(\sum_{i=1}^n m_i)$, and $Y_{ij}(\cdot)=Y_i(\cdot, T_{ij})$. Fix temporarily $s,s'\in \mathcal{S}$ and define $A_{i}(T)=Y_i(s,T)Y_i(s',T)$; then $A_i$ is iid as the random variable $ A$ defined by $A(T) = Y(s,T)Y(s',T)$. Let $\{T_d: d=1,\ldots, D\}$ be a set of unique $T_{ij}$'s for all $i$ and $j$ in increasing order such that $T_1 < T_2 < \ldots < T_D$ with $T_0 = 0$ and $T_D=1$. $D$ denotes the total number of unique $T_{ij}$'s. Due to the sparse assumption of the longitudinal design ($m_i<\infty$ for all $i=1, \ldots, n$) we obtain that $D$ diverges with the sample size $n$. It follows that
\begin{eqnarray*}
\widehat{\Sigma}(s,s')& =& \frac{1}{\sum_{i=1}^n m_i}  \sum_{d=1}^D  \sum_{i=1}^{n}\sum_{j=1}^{m_i}A_i(T_d)I(T_{ij} \in (T_{d-1},T_{d}])\\
&=& \frac{1}{D}  \sum_{d=1}^D  \frac{D}{M_1} \sum_{i=1}^{n}\sum_{j=1}^{m_i}A_i(T_d)I(T_{ij} \in (T_{d-1},T_{d}])\\
&=& \frac{1}{D}  \sum_{d=1}^D   V_d;
\end{eqnarray*}
the first equality is obtained by using a different way of counting the summands of $\widehat \Sigma(s,s')$, and the second one by multiplying and diving by $D$. Let $M_1 = \sum_{i=1}^{n} m_i$, $M_2 = \sum_{i=1}^{n} m_i^2$ and $P_d = P({\rm T} \in (T_{d-1}, T_d])$, where by an abuse of notation ${\rm T}$ denotes the random variable with sampling distribution $g(T)$. Here we defined $V_d = (D/M_{1}) \sum_{i=1}^{n}\sum_{j=1}^{m_i}A_i(T_d)I(T_{ij} \in (T_{d-1},T_{d}])$; $V_d$ depends on $s, s'$, though the dependence on $s$ and $s'$ is suppressed. This $\widehat{\Sigma}(s,s') = \overline{V} = D^{-1}\sum_{d=1}^{D} V_d$. $V_d$'s are correlated over $d$; thus to show that `$\widehat{\Sigma}(s,s')$ is consistent' it is sufficient to show that the average of dependent variables $V_d$'s is consistent.

Take the latter problem and study first the covariance of $V_{d}$ and $V_{d'}$. We have: 
\begin{align} \nonumber
	\text{cov}&(V_d, V_{d'}) = \text{E}[V_dV_{d'}] - \text{E}[V_d]\text{E}[V_{d'}] \\ \nonumber
	&= \begin{cases}
		\Big\{\dfrac{D^2}{M_1}\Big\}\Big\{P_d\cdot \text{E}[A(T)A(T')| T, T'])\Big\}\\[0.8em]
		\qquad \qquad \qquad - \text{ } \Big\{\dfrac{D^2M_2}{M_1^2}\Big\} \big\{P_d \cdot c((s,T_d),(s',T_d))\big\}^2 \qquad \text{ for } d = d' \\\\
		\Big\{\dfrac{M_2D^2}{M_1^2} - \dfrac{D^2}{M_1}\Big\} \Big\{P_d  P_{d'} \cdot \text{E}[A(T)A(T')| T, T'] \Big\}\\[0.8em]
		\qquad \qquad \qquad - \text{ } \Big\{\dfrac{M_2D^2}{M_1^2}\Big\} \Big\{ P_d  P_{d'} \cdot c((s,T_d),(s',T_d)) \cdot c((s,T_{d'}),(s',T_{d'})) \Big\} \qquad \text{ for } d \neq d' .
	\end{cases}
\end{align}

For simplicity in notation, denote the variance of $V_d$ with $\sigma_d^2$, and the covariance of $V_d$ and $V_{d'}$ with $\sigma_{dd'}$. Under the assumptions \ref{assmp: square int X}-\ref{assmp: E[A(T)A(T')|T,T']}, it holds that $\text{E}(V_d)$, $\sigma_d^2$, and $\sigma_{dd'}$ are finite. Following \textit{Theorem 5.3 \citep[p.208]{boos2013essential}}, we show the consistency of $\overline{V}$ by showing that the following converges to $0$ in probability as $D$ diverges:

\begin{align}\nonumber
	\dfrac{1}{D^2} \Big[\sum_{d=1}^{D} \sigma_d^2 + \sum_{d=1}^{D}\sum_{d'\neq d} \sigma_{dd'}\Big] =& \Bigg[\dfrac{1}{M_1} \text{E}_T[\text{E}[A(T)^2|T]] \Bigg] + \Bigg[\Big\{\dfrac{M_2}{M_1^2} - \dfrac{1}{M_1}\Big\} \text{E}_T[ \text{E}_{T'}[\text{E}[A(T)A(T')|T,T']]] \Bigg]\\ \nonumber 
	& \qquad - \Bigg[\Big\{\dfrac{M_2}{M_1^2} - \dfrac{1}{M_1}\Big\} \sum_{d=1}^{D}P^2_d \cdot \text{E}[A(T)^2|T=T_d] \Bigg] - \dfrac{M_2}{M_1^2} \{\Sigma(s,s')\}^2
\end{align}\\
As integration of a continuous function in a compact interval is finite, under the assumptions \ref{assmp: reg con g(T)} and \ref{assmp: E[A(T)A(T')|T,T']} $\text{E}_T[\text{E}[A(T)^2|T]]$, $\text{E}_T[ \text{E}_{T'}[\text{E}[A(T)A(T')|T,T']]]$, and $\Sigma(s,s')$ are finite. For example  $\text{E}_T[\text{E}[A(T)^2|T]] = \int_{\mathcal{T}} g(T) \text{E}[A(T)^2|T] dT \leq \text{sup}_{T} g(T) \int_{\mathcal{T}} \text{E}[A(T)^2|T] dT$ is finite because (i) $\text{sup}_{T} g(T)$ is finite by the assumption \ref{assmp: reg con g(T)} and (ii) $\int_{\mathcal{T}} \text{E}[A(T)^2|T] dT$ is finite because $\text{E}[A(T)^2|T]$ is continuous and finite in $\mathcal{T}$ by the assumptions \ref{assmp: reg con g(T)} and \ref{assmp: E[A(T)A(T')|T,T']}. With the same argument we can show that $\sum_{d=1}^{D}P^2_d \cdot \text{E}[A(T)^2|T=T_d] < \text{sup}_{d}P_d \text{E}_T[\text{E}[A(T)^2|T]]$ is also finite. Here we use the results that $M_1/M_2 = O(1)$ and each term diverges to $\infty$ as $D$ diverges (since $D \leq M_1$). 
It implies that
\begin{equation*}
\dfrac{1}{D^2} \Big[\sum_{d=1}^{D} \sigma_d^2 + \sum_{d=1}^{D}\sum_{d'\neq d} \sigma_{dd'}\Big] \xrightarrow{\text{  p  }} 0 \text{ as } D \rightarrow \infty.
\end{equation*} 
Using \textit{Theorem 5.3 \citep{boos2013essential}}, we obtain that $\widehat{\Sigma}(s,s') = \overline{V}= D^{-1}\sum_{d=1}^{D} V_d$ converges in probability to $D^{-1}\sum_{d=1}^{D} \text{E}[V_d] = \sum_{d=1}^{D} P_d \cdot c((s,T_d),(s',T_d))$ as $n$ diverges; where the latter expression is equal to $ \Sigma(s,s')$.\\


\noindent  \emph{Hilbert-Schmidt norm consistency}\\

Let $R_{ijdv} = <Y_{ij}, e_v>Y_{ij} I(T_{ij}\in (T_{d-1},T_d]) = <Y_{i}(\cdot, T_d), e_v>Y_{i}(\cdot, T_d) I(T_{ij}\in (T_{d-1},T_d])$, where $\{e_v: v \geq 1\}$ is any orthonormal basis. Define the sample covariance operator associated with $\widehat{\Sigma}(s,s')$ as follows:
\begin{align}\nonumber
	H(x) &= D^{-1}\sum_{d=1}^{D} \dfrac{D}{M_1} \sum_{i=1}^{n}\sum_{j=1}^{m_i} <Y_{ij}, x>Y_{ij} I(T_{ij}\in (T_{d-1},T_d])   \\ \nonumber
	& = D^{-1}\sum_{d=1}^{D} \dfrac{D}{M_1} \sum_{i=1}^{n}\sum_{j=1}^{m_i} <Y_i(\cdot,T_d), x>Y_{i}(\cdot,T_d) I(T_{ij}\in (T_{d-1},T_d]), \qquad x \in L^2.
\end{align}

\noindent Under the assumptions \ref{assmp: square int X}, \ref{assmp: reg con g(T)}, and \ref{assmp: 4th moment Y}, we have that the following sequence of ine/equalities are true:
{\allowdisplaybreaks\begin{align}\nonumber
		&\text{E}\Big[  \big\| \widehat{\Sigma}(s,s')-\text{E}[\widehat{\Sigma}(s,s')] \big\|^2_s\Big] = \text{E} \Big[  \big\|H - \text{E}[H]  \big\|^2_s   \Big] =  \text{E} \Big[ \sum_{v=1}^{\infty}  \big\|H(e_v) - \text{E}[H(e_v)]  \big\|^2_s   \Big]  \\ \nonumber
		& = \text{E}\Bigg[ \sum_{v=1}^{\infty}   \Big\| D^{-1} \sum_{d=1}^{D} \dfrac{D}{M_1} \sum_{i=1}^{n}\sum_{j=1}^{m_i} \Big\{  <Y_{ij}, e_v>Y_{ij} I(T_{ij}\in (T_{d-1},T_d]) \\ \nonumber
		& \qquad \qquad \qquad \qquad \qquad \qquad \qquad \qquad \qquad \qquad\qquad   - \text{E}[ <Y_{ij}, e_v>Y_{ij} I(T_{ij}\in (T_{d-1},T_d]) ] \Big\} \Big\|^2_s   \Bigg]\\ \nonumber
		& \text{(using Equation (2.2) from \citet[p.22]{horvath2012inference})}\\ \nonumber
		& = \text{E} \Big[ \sum_{v=1}^{\infty}   \Big\| D^{-1} \sum_{d=1}^{D} \dfrac{D}{M_1} \sum_{i=1}^{n}\sum_{j=1}^{m_i} \Big\{ R_{ijdv}-\text{E}[R_{ijdv}] \Big\} \Big\|^2_s  \Big]\\ \nonumber
		& = \text{E} \Big[  \sum_{v=1}^{\infty}  D^{-2} \sum_{d=1}^{D} \sum_{d'=1}^{D} \dfrac{D^2}{M_1^2} 
		<\sum_{i}^{n}\sum_{j}^{m_i} \big\{ R_{ijdv}-\text{E}[R_{ijdv}] \big\}, \sum_{i'}^{n} \sum_{j'}^{m_{i'}} \big\{ R_{i'j'd'v}-\text{E}[R_{i'j'd'v}] \big\}>   \Big]\\ \nonumber
		& =  D^{-2} \sum_{d=1}^{D} \sum_{d'=1}^{D} \dfrac{D^2}{M_1^2} \Bigg[ \sum_{i}^{n}\sum_{i'}^{n}\sum_{j}^{m_i}\sum_{j'}^{m_{i'}} \sum_{v=1}^{\infty} \text{E}\Big\{ <R_{ijdv}-\text{E}[R_{ijdv}], R_{i'j'd'v}-\text{E}[R_{i'j'd'v}]>\Big\}   \Bigg] \\ 
		& \leq M_1^{-2} \sum_{d=1}^{D} \sum_{d'=1}^{D} \Bigg| \sum_{i}^{n}\sum_{i'}^{n}\sum_{j}^{m_i}\sum_{j'}^{m_{i'}} \sum_{v=1}^{\infty} \text{E}\Big\{ <R_{ijdv}-\text{E}[R_{ijdv}], R_{i'j'd'v}-\text{E}[R_{i'j'd'v}]>\Big\} \Bigg |\\ \nonumber
		& \leq M_1^{-2} \sum_{d=1}^{D} \sum_{d'=1}^{D} \sum_{i}^{n}\sum_{i'}^{n}\sum_{j}^{m_i}\sum_{j'}^{m_{i'}} \sum_{v=1}^{\infty} \Bigg| \Big\{ \text{E}[<R_{ijdv}, R_{i'j'd'v}>] - < \text{E}[R_{ijdv}]  , \text{ E}[R_{i'j'd'v}]  > \Big\} \Bigg|\\ \nonumber
		& \leq \Big\{ \dfrac{(M_2-M_1)\cdot \text{sup}_d P_d }{M_1^2} + \dfrac{1}{M_1} + \dfrac{2(M_2-M_1)}{M_1^2} + \dfrac{1}{M_1} \Big\} \cdot \text{ E}_T \Bigg[ \text{E}\Big[ \big\|Y(\cdot, T_d) \big\|^4   \Big] \Bigg] \longrightarrow 0, 
	\end{align}
}
as $n$ diverges. It implies that $ \text{ E} \Big[ \big\| \widehat{\Sigma}(s,s') - \text{E}[\widehat{\Sigma}(s,s')]  \big\|^2_s   \Big] =  \text{ E} \Big[ \big\| \widehat{\Sigma}(s,s') - \Sigma(s,s')  \big\|^2_s   \Big]$ converges to $0$ as $n$ diverges, and the Hilbert-Schmidt norm consistency, $\big\| \widehat{\Sigma}(s,s')- \Sigma(s,s') \big\|_s \xrightarrow{\text{  p  }} 0$, is implied by Markov inequality.\\

\subsection*{Proof for Theorem \ref{th: consistency of xi and G}}
First we show that for each $k$ 
\begin{equation} \label{Appeq: uniform convergence of xi}
	\text{sup}_{j} |\widetilde{\xi}_{W,ijk} - \xi_{W, ijk}| \xrightarrow{\text{  p  }} 0,
\end{equation}
as $n$ diverges. 

We have that the following sequence of inequalities hold:	
{\allowdisplaybreaks 
	\begin{align}  \nonumber
		\text{sup}_j  |\widetilde{\xi}_{W,ijk} - \xi_{W, ijk}| & = \text{sup}_j \Big| \int Y_i(s,T_{ij}) \widehat{\phi}_k(s) ds - \int Y_i(s,T_{ij}) \phi_k(s) ds \Big| \\\nonumber
		& = \text{sup}_j \Big| \int Y_i(s,T_{ij}) \{ \widehat{\phi}_k(s) - \phi_k(s) \} ds \Big|\\\nonumber
		& \leq \text{sup}_j  \int \Big| Y_i(s,T_{ij}) \Big| \Big| \{ \widehat{\phi}_k(s) - \phi_k(s) \}\Big|  ds \\\nonumber
		& \leq \text{sup}_j  \text{sup}_{s \in [0,1]} \Big|Y_i(s,T_{ij}) \Big| \int \Big|\{ \widehat{\phi}_k(s) - \phi_k(s)\Big| \}  ds\\\nonumber
		& \leq \text{sup}_{j, s} \Big| Y_{i}(s, T_{ij})\Big| \times \Big\{ \int \{ \widehat{\phi}_k(s) - \phi_k(s) \}^2  ds\Big\}^{1/2} \qquad \text{(by Cauchy-Schwartz ineq.)} \\
		& = \text{sup}_{j, s} \Big|Y_{i}(s, T_{ij})\Big| \times \|\widehat\phi_k(\cdot) - \phi_k(\cdot) \|_s, \label{Appeq: score convergence}
	\end{align}
} %
where $\text{sup}_{j, s} \Big|Y_{i}(s, T_{ij})\Big|$ is absolutely bounded almost surely under the assumption \ref{assmp: Y asymptotically bounded} and $\|\widehat\phi_k(\cdot) - \phi_k(\cdot) \|_s$ converges to $0$ as shown in Corollary \ref{cor: consistency of eigencomponents of mFPCA} that is proved in the Supplementary Material. This concludes the first part of the proof.
\\
\\
\noindent Recall that $G_{k}(T,T') = \text{cov}\{\xi_{il}(T), \xi_{ik}(T')\}$ is the true covariance. It is already shown in \citet{yao2005functional} the uniform consistency of its local linear estimator, $\widehat{G}_{W,k}(T,T')$, when $\widehat{G}_{W,k}(T,T')$ is obtained with $\xi_{W, ijk}$'s; specifically, $\widehat{G}_{W,k}(T, T')$ is obtained by the local linear smoothing of $\{\widetilde{G}_{W,ik}(T_{ij}, T_{ij'})=\xi_{W, ijk}\xi_{W,ij'k}: j \neq j'\}$. Here we show a similar result when the local linear estimator is obtained with $\widetilde{\xi}_{W,ijk}$ instead of $\xi_{W, ijk}$.
		
The sample covariance of $\widetilde{\xi}_{W,ijk}$'s is as follows:
\begin{align*}
	\widetilde{G}_{ik}(T_{ij}, T_{ij'}) &= \widetilde{\xi}_{W,ijk} \widetilde{\xi}_{W,ij'k}\\
	& = \{(\widetilde{\xi}_{W,ijk} - \xi_{W,ijk}) + \xi_{W,ijk})\}
	\times \{(\widetilde{\xi}_{W,ij'k} - \xi_{W,ij'k}) + \xi_{W,ij'k}\}\\
	& = \widetilde{G}_{W,ik}(T_{ij}, T_{ij'})+(\widetilde{\xi}_{W,ij'k} - \xi_{W,ij'k})(\widetilde{\xi}_{W,ijk} - \xi_{W,ijk}) \\
	& \qquad \qquad + \xi_{W,ij'k}(\widetilde{\xi}_{W,ijk} - \xi_{W,ijk}) + \xi_{W,ijk}(\widetilde{\xi}_{W,ij'k} - \xi_{W,ij'k})
\end{align*}
By the uniform consistency of $\widetilde{\xi}_{W,ijk}$ given in Equation (\ref{Appeq: uniform convergence of xi}), the local linear estimator, $\widehat{G}_{k}(T,T')$, obtained by smoothing $\{ (T_{ij}, T_{ij'}), \widetilde{G}_{ik}(T_{ij}, T_{ij'}) :i=1, \ldots,n, j\neq j'\}$ is asymptotically equivalent to the local linear estimator, $\widehat{G}_{W,k}(T,T')$, obtained by smoothing $\widetilde{G}_{W,ik}(T_{ij}, T_{ij'})$'s. Furthermore as $\widehat{G}_{W,k}(T,T')$ is shown to be a uniformly consistent estimator \citep{yao2005functional}, the uniform consistency of $\widehat{G}_{k}(T,T')$ follows. Furthermore the Hilbert-Schmidt norm consistency is implied by the uniform consistency. \\

\subsection*{Proof for Conjecture \ref{th: consistency of predicted trajectories}}

In the following we prove the consistency of the predicted trajectory, $\widehat{Y}_i(s,T)$, for the case when $\xi_{W, ijk}$'s are uncorrelated over $k$. This is basis of our intuition for Conjecture \ref{th: consistency of predicted trajectories}. 
		
To show the consistency of the predicted trajectory, $\widehat{Y}_i(s,T)=\sum_{k=1}^{K} \widehat{\xi}_{ik}(T) \widehat{\phi}_k(s)=\sum_{k=1}^{K} \sum_{l=1}^{L_k} \widehat{\zeta}_{ikl} \widehat{\psi}_{kl}(T) \widehat{\phi}_k(s)$, we first need to show that the truncated true trajectories, $\widetilde{Y}^{KL}_{i}(s,T) = \sum_{k=1}^{K} \sum_{l=1}^{L_k} \zeta_{ikl} \psi_{kl}(T) \phi_k(s)$, are consistent estimators of the corresponding (not-truncated) true trajectories, $\widetilde{Y}_{i}(s,T) = \sum_{k=1}^{\infty} \sum_{l=1}^{\infty} \zeta_{ikl} \psi_{kl}(T) \phi_k(s)$.

Recall that $\{\phi_k(s)\}_k$ is set as the eigenbasis of the marginal covariance $\Sigma(s,s')$. It follows that for as $K\rightarrow \infty $ we have that $\sum_{k=K+1}^{\infty}\lambda_k \phi_k(s) \phi_k(s')$ converges to $0$ uniformly over $s$ and $s'$ by Mercer's theorem. And it implies that $\sum_{k=K+1}^{\infty} \lambda_k \phi_k(s)^2$ also converges to $0$ uniformly over $s$, and furthermore that

\allowdisplaybreaks{\begin{align} \nonumber
	\text{sup}_{s} & \sum_{k=K+1}^{\infty}  \lambda_k \phi_k(s)^2 = \text{sup}_{s} \text{var}\Big[\sum_{k=K+1}^{\infty} \xi_{W,ijk} \phi_k(s) \Big] \\ \nonumber
	& = \text{sup}_{s} \text{var}\Big[\sum_{k=K+1}^{\infty} \big\{  \xi_{ik}(T_{ij})+e_{ijk}\big\} \phi_k(s) \Big] \\ \nonumber 
	& = \text{sup}_{s} \text{var}\Big[\sum_{k=K+1}^{\infty} \sum_{l=1}^{\infty} \zeta_{ikl} \psi_{kl}(T_{ij}) \phi_k(s)\Big] + \text{var}\Big[\sum_{k=K+1}^{\infty}  e_{ijk}\phi_k(s)\Big] \\ \nonumber
	&=\text{sup}_{s} \text{var} \Big[ \text{E} \Big\{ \sum_{k=K+1}^{\infty} \sum_{l=1}^{\infty}\zeta_{ikl} \psi_{kl}(T_{ij})\phi_k(s) \big| \boldsymbol{\xi}_{W,ik} \Big\}\Big] \\ \nonumber
	& \qquad \qquad \qquad + \text{E} \Big[ \text{var} \Big\{\sum_{k=K+1}^{\infty} \sum_{l=1}^{\infty}\zeta_{ikl} \psi_{kl}(T_{ij})\phi_k(s) \big| \boldsymbol{\xi}_{W,ik} \Big\}\Big] + \sum_{k=K+1}^{\infty} \sigma^{2}_{e,k}\phi^{2}_k(s) \\ \nonumber
	& \qquad \text{(By total law of variance)} \\ \nonumber
	& =  \text{sup}_{s} \text{E} \Big[\Big\{ \text{E} \Big(\sum_{k=K+1}^{\infty} \sum_{l=1}^{\infty}\zeta_{ikl} \psi_{kl}(T_{ij}) \phi_k(s) \big| \boldsymbol{\xi}_{W,ik} \Big) \Big\}^2\Big] \\ \label{Appeq: total law of variance}
	& \qquad \qquad+  \text{E} \Big[ \text{var} \Big\{\sum_{k=K+1}^{\infty} \sum_{l=1}^{\infty}\zeta_{ikl} \psi_{kl}(T_{ij})\big\} \phi_k(s) \big| \boldsymbol{\xi}_{W,ik}\Big\}\Big] + \sum_{k=K+1}^{\infty}  \sigma^{2}_{e,k}\phi^{2}_k(s) \longrightarrow 0,
\end{align}
}
as $K$ diverges. Because all terms in Equation (\ref{Appeq: total law of variance}) are non-negative and the summation of those terms converges to $0$, it is implied that each term also converges to $0$ as $K$ diverges. And by Markov inequality, $ \text{E} \Big\{\sum_{k=K+1}^{\infty} \sum_{l=1}^{\infty}\zeta_{ikl} \psi_{kl}(T_{ij}) \phi_k(s) \big| \boldsymbol{\xi}_{W,ik} \Big\}$ converges to $0$ as $K$ diverges. Notice that 
\begin{align}\nonumber
	\text{E} \Big\{\sum_{k=K+1}^{\infty} \sum_{l=1}^{\infty}\zeta_{ikl} & \psi_{kl}(T_{ij}) \phi_k(s) \big| \boldsymbol{\xi}_{W,ik} \Big\}  = \sum_{k=K+1}^{\infty} \sum_{l=1}^{\infty} \text{E}\big[\zeta_{ikl} \big| \boldsymbol{\xi}_{W,ik} \big] \psi_{kl}(T_{ij}) \phi_k(s) \\ \nonumber
	& = \sum_{k=K+1}^{\infty} \sum_{l=1}^{\infty} \widetilde{\zeta}_{ikl} \psi_{kl}(T_{ij}) \phi_k(s) \\ \nonumber
	& = \widetilde{Y}_{i}(s,T_{ij}) - \sum_{k=1}^{K} \widetilde{\xi}_{ik}(T_{ij}) \phi_k(s) \longrightarrow 0 \text{ as } K \longrightarrow \infty,
\end{align}
where $\widetilde{\xi}_{ik}(T)=\sum_{l=1}^{\infty} \widetilde{\zeta}_{ikl} \psi_{kl}$ is the full target trajectory that we want to predict in the second FPCA step. 
		
Let $\widetilde{\xi}^{L}_{ik}(T) = \sum_{l=1}^{L_k} \widetilde{\zeta}_{ikl} \psi_{kl}$ be the truncated true trajectory of $\widetilde{\xi}_{ik}(T)$. By Lemma 3 of \citet{yao2005functional}, $\text{sup}_T \; \text{E}[\widetilde{\xi}^{L}_{ik}(T) - \widetilde{\xi}_{ik}(T)]^2 \longrightarrow 0$, and by Markov inequality it implies that for each $k$ $\widetilde{\xi}^{L}_{ik}(T)$ converges to $\widetilde{\xi}_{ik}(T)$ in probability as $L_K$ diverges. Thus by combining two results, (1) $\sum_{k=1}^{K} \widetilde{\xi}_{ik}(T_{ij}) \phi_k(s)$ converges to $\widetilde{Y}_{i}(s,T_{ij})$ in probability as $K$ diverges and (2) for each $k$ $\widetilde{\xi}^{L}_{ik}$ converges to $\widetilde{\xi}_{ik}(T)$ in probability as $L_k$ diverges, we show that the truncated target trajectories, $\widetilde{Y}^{KL}_{i}(s,T)$'s, converge to the corresponding full target trajectories, $\widetilde{Y}_{i}(s,T)$'s.
		
Finally we show that the predicted trajectory, $\widehat{Y}_{i}(s,T)$, is a consistent estimator of the full target trajectory, $\widetilde{Y}_{i}(s,T)$. Note that $|\widehat{Y}_{i}(s,T) - \widetilde{Y}^{KL}_{i}(s,T)| \leq |\widehat{Y}_{i}(s,T) - \widetilde{Y}^{KL}_{i}(s,T)| + |\widetilde{Y}^{KL}_{i}(s,T) - \widetilde{Y}_{i}(s,T)|$. We already show that the second term converges to $0$ as $K$ and $L_k$'s diverge. And by Slutsky's theorem and the consistency of the estimators of all model components, the first term, $|\widehat{Y}_{i}(s,T) - \widetilde{Y}^{KL}_{i}(s,T)|$, converges to $0$ in probability as $n$ diverges. Thus for each $(s,T) \in \mathcal{S} \times \mathcal{T}$,
\begin{equation}
	\widehat{Y}_{i}(s,T) = \sum_{k=1}^{K} \sum_{l=1}^{L_k} \widehat{\zeta}_{ikl} \widehat{\psi}_{kl}(T)\widehat{\phi}_k(s) \xrightarrow{\text{  p  }} \sum_{k=1}^{\infty} \sum_{l=1}^{\infty} \widetilde{\zeta}_{ikl} \psi_{kl}(T)\phi_k(s),
\end{equation}
as $n$, $K$ and $L_k$'s diverge.

	\section*{Acknowledgement} \label{sec:Ack}
	Staicu's research was supported by NSF grant number DMS 1454942
	and NIH grant R01 NS085211. We thank Daniel Reich and Peter Calabresi for the DTI tractography data.
	
	\section*{Supplementary Material} \label{sec:SM}
	
	Detailed proof of the theoretical results, additional numerical investigations and data analysis results are included in a supplementary material that is available online.


\fontsize{12pt}{12pt}
		
\clearpage
		
\clearpage
\bibliography{reference}

\begin{thebibliography}{}

\bibitem[Alexander et~al., 2007]{alexander2007diffusion}
Alexander, A.~L., Lee, J.~E., Lazar, M., and Field, A.~S. (2007).
\newblock Diffusion tensor imaging of the brain.
\newblock {\em Neurotherapeutics}, 4(3):316--329.

\bibitem[Baladandayuthapani et~al., 2008]{baladandayuthapani2008bayesian}
Baladandayuthapani, V., Mallick, B.~K., Young~Hong, M., Lupton, J.~R., Turner,
  N.~D., and Carroll, R.~J. (2008).
\newblock Bayesian hierarchical spatially correlated functional data analysis
  with application to colon carcinogenesis.
\newblock {\em Biometrics}, 64(1):64--73.

\bibitem[Basser et~al., 1994]{basser1994mr}
Basser, P.~J., Mattiello, J., and LeBihan, D. (1994).
\newblock Mr diffusion tensor spectroscopy and imaging.
\newblock {\em Biophysical journal}, 66(1):259.

\bibitem[Basser et~al., 2000]{basser2000vivo}
Basser, P.~J., Pajevic, S., Pierpaoli, C., Duda, J., and Aldroubi, A. (2000).
\newblock In vivo fiber tractography using dt-mri data.
\newblock {\em Magnetic resonance in medicine}, 44(4):625--632.

\bibitem[Basser and Pierpaoli, 2011]{basser2011microstructural}
Basser, P.~J. and Pierpaoli, C. (2011).
\newblock Microstructural and physiological features of tissues elucidated by
  quantitative-diffusion-tensor mri.
\newblock {\em Journal of magnetic resonance}, 213(2):560--570.

\bibitem[Boos and Stefanski, 2013]{boos2013essential}
Boos, D.~D. and Stefanski, L. (2013).
\newblock {\em Essential Statistical Inference}.
\newblock Springer.

\bibitem[Bosq, 2000]{bosq2000linear}
Bosq, D. (2000).
\newblock {\em Linear processes in function spaces: theory and applications},
  volume 149.
\newblock Springer.

\bibitem[Cardot et~al., 2003]{cardot2003testing}
Cardot, H., Ferraty, F., Mas, A., and Sarda, P. (2003).
\newblock Testing hypotheses in the functional linear model.
\newblock {\em Scandinavian Journal of Statistics}, 30(1):241--255.

\bibitem[Cardot et~al., 2004]{cardot2004testing}
Cardot, H., Goia, A., and Sarda, P. (2004).
\newblock Testing for no effect in functional linear regression models, some
  computational approaches.
\newblock {\em Communications in Statistics-Simulation and Computation},
  33(1):179--199.

\bibitem[Chang et~al., 2015]{shiny}
Chang, W., Cheng, J., Allaire, J., Xie, Y., and McPherson, J. (2015).
\newblock {\em shiny: Web Application Framework for R}.
\newblock R package version 0.12.1.

\bibitem[Chen and M{\"u}ller, 2012]{chen2013repeated}
Chen, K. and M{\"u}ller, H.-G. (2012).
\newblock Modeling repeated functional observations.
\newblock {\em Journal of the American Statistical Association},
  107(500):1599--1609.

\bibitem[{Ciprian Crainiceanu} et~al., 2014]{Rrefundpackage}
{Ciprian Crainiceanu}, {Philip Reiss}, {Jeff Goldsmith}, {Lei Huang}, {Lan
  Huo}, and {Fabian Scheipl} (2014).
\newblock {\em refund: Regression with Functional Data}.
\newblock R package version 0.1-11.

\bibitem[Crainiceanu et~al., 2011]{crainiceanu2011statistical}
Crainiceanu, C.~M., Staicu, A.-M., Ray, S., and Punjabi, N. (2011).
\newblock Statistical inference on the difference in the means of two
  correlated functional processes: an application to sleep eeg power spectra.
\newblock {\em Johns Hopkins University, Dept. of Biostatistics Working
  Papers}, page 225.

\bibitem[Di et~al., 2009]{di2009multilevel}
Di, C.-Z., Crainiceanu, C.~M., Caffo, B.~S., and Punjabi, N.~M. (2009).
\newblock Multilevel functional principal component analysis.
\newblock {\em The annals of applied statistics}, 3(1):458.

\bibitem[Goldsmith et~al., 2011]{goldsmith2011penalized}
Goldsmith, J., Bobb, J., Crainiceanu, C.~M., Caffo, B., and Reich, D. (2011).
\newblock Penalized functional regression.
\newblock {\em Journal of Computational and Graphical Statistics}, 20(4).

\bibitem[Goldsmith et~al., 2014]{goldsmith2014generalized}
Goldsmith, J., Zipunnikov, V., and Schrack, J. (2014).
\newblock Generalized multilevel functional-on-scalar regression and principal
  component analysis.

\bibitem[Greven et~al., 2010]{greven2010}
Greven, S., Crainiceanu, C., Caffo, B., and Reich, D. (2010).
\newblock Longitudinal functional principal component analysis.
\newblock {\em Electronic Journal of Statistics}, 4:1022--1054.

\bibitem[Gromenko and Kokoszka, 2013]{gromenko2013nonparametric}
Gromenko, O. and Kokoszka, P. (2013).
\newblock Nonparametric inference in small data sets of spatially indexed
  curves with application to ionospheric trend determination.
\newblock {\em Computational Statistics \& Data Analysis}, 59:82--94.

\bibitem[Gromenko et~al., 2012]{gromenko2012}
Gromenko, O., Kokoszka, P., Zhu, L., and Sojka, J. (2012).
\newblock Estimation and testing for spatially indexed curves with application
  to ionospheric and magnetic field trends.
\newblock {\em The Annals of Applied Statistics}, 6(2):669--696.

\bibitem[Hastie et~al., 2009]{hastie2009elements}
Hastie, T., Tibshirani, R., Friedman, J., Hastie, T., Friedman, J., and
  Tibshirani, R. (2009).
\newblock {\em The elements of statistical learning}, volume~2.
\newblock Springer.

\bibitem[Horv{\'a}th and Kokoszka, 2012]{horvath2012inference}
Horv{\'a}th, L. and Kokoszka, P. (2012).
\newblock {\em Inference for functional data with applications}, volume 200.
\newblock Springer.

\bibitem[James et~al., 2000]{james2000principal}
James, G.~M., Hastie, T.~J., and Sugar, C.~A. (2000).
\newblock Principal component models for sparse functional data.
\newblock {\em Biometrika}, 87(3):587--602.

\bibitem[Jiang and Wang, 2010]{jiang2010covariate}
Jiang, C.-R. and Wang, J.-L. (2010).
\newblock Covariate adjusted functional principal components analysis for
  longitudinal data.
\newblock {\em The Annals of Statistics}, pages 1194--1226.

\bibitem[Li and Guan, 2014]{li2014fpca}
Li, Y. and Guan, Y. (2014).
\newblock Functional principal component analysis of spatio-temporal point
  processes with applications in disease surveillance.
\newblock {\em Journal of the American Statistical Association}, 0(ja):null.

\bibitem[Marx and Eilers, 2005]{marx2005multidimensional}
Marx, B.~D. and Eilers, P.~H. (2005).
\newblock Multidimensional penalized signal regression.
\newblock {\em Technometrics}, 47(1):13--22.

\bibitem[MATLAB, 2014]{MATLAB:2014}
MATLAB (2014).
\newblock {\em version 8.4.0 (R2014b)}.
\newblock The MathWorks Inc., Natick, Massachusetts.

\bibitem[Morris and Carroll, 2006]{morris2006wavelet}
Morris, J.~S. and Carroll, R.~J. (2006).
\newblock Wavelet-based functional mixed models.
\newblock {\em Journal of the Royal Statistical Society: Series B (Statistical
  Methodology)}, 68(2):179--199.

\bibitem[Morris et~al., 2003]{morris2003wavelet}
Morris, J.~S., Vannucci, M., Brown, P.~J., and Carroll, R.~J. (2003).
\newblock Wavelet-based nonparametric modeling of hierarchical functions in
  colon carcinogenesis.
\newblock {\em Journal of the American Statistical Association},
  98(463):573--583.

\bibitem[Park et~al., 2015]{park2015fixed}
Park, S.~Y., Staicu, A.-M., Xiao, L., and Crainiceanu, C.~M. (2015).
\newblock Simple fixed effects inference for complex functional models.
\newblock Manuscript submitted.

\bibitem[Peng and Paul, 2009]{peng2009geometric}
Peng, J. and Paul, D. (2009).
\newblock A geometric approach to maximum likelihood estimation of the
  functional principal components from sparse longitudinal data.
\newblock {\em Journal of Computational and Graphical Statistics}, 18(4).

\bibitem[Pomann et~al., 2013]{pomann2013two}
Pomann, G.-M., Staicu, A.-M., and Ghosh, S. (2013).
\newblock Two sample hypothesis testing for functional data.

\bibitem[{R Core Team}, 2014]{Rsoftware}
{R Core Team} (2014).
\newblock {\em R: A Language and Environment for Statistical Computing}.
\newblock R Foundation for Statistical Computing, Vienna, Austria.

\bibitem[Staicu et~al., 2010]{staicu2010fast}
Staicu, A.-M., Crainiceanu, C.~M., and Carroll, R.~J. (2010).
\newblock Fast methods for spatially correlated multilevel functional data.
\newblock {\em Biostatistics}, 11(2):177--194.

\bibitem[Staicu et~al., 2012]{staicu2012modeling}
Staicu, A.-M., Crainiceanu, C.~M., Reich, D.~S., and Ruppert, D. (2012).
\newblock Modeling functional data with spatially heterogeneous shape
  characteristics.
\newblock {\em Biometrics}, 68(2):331--343.

\bibitem[Staniswalis and Lee, 1998]{staniswalis1998nonparametric}
Staniswalis, J.~G. and Lee, J.~J. (1998).
\newblock Nonparametric regression analysis of longitudinal data.
\newblock {\em Journal of the American Statistical Association},
  93(444):1403--1418.

\bibitem[Wood, 2006]{wood2006a}
Wood, S.~N. (2006).
\newblock Low-rank scale-invariant tensor product smooths for generalized
  additive mixed models.
\newblock {\em Biometrics}, 62(4):1025--1036.

\bibitem[Wood, 2011]{wood2011faststable}
Wood, S.~N. (2011).
\newblock Fast stable restricted maximum likelihood and marginal likelihood
  estimation of semiparametric generalized linear models.
\newblock {\em Journal of the Royal Statistical Society: Series B (Statistical
  Methodology)}, 73(1):3--36.

\bibitem[Xiao et~al., 2013]{xiao2013fast}
Xiao, L., Li, Y., and Ruppert, D. (2013).
\newblock Fast bivariate p-splines: the sandwich smoother.
\newblock {\em Journal of the Royal Statistical Society: Series B (Statistical
  Methodology)}, 75(3):577--599.

\bibitem[Yao et~al., 2005]{yao2005functional}
Yao, F., M{\"u}ller, H.-G., and Wang, J.-L. (2005).
\newblock Functional data analysis for sparse longitudinal data.
\newblock {\em Journal of the American Statistical Association},
  100(470):577--590.

\bibitem[Zhang and Chen, 2007]{zhang2007statistical}
Zhang, J.-T. and Chen, J.~W. (2007).
\newblock Statistical inferences for functional data.
\newblock {\em The Annals of Statistics}, 35(3):1052--1079.

\end{thebibliography}
\bibliographystyle{apalike}

\end{document}